\documentclass[12pt]{TD-CJS}

\usepackage{latexsym}
\usepackage{amsmath}
\usepackage{amsfonts}
\usepackage{amssymb}
\usepackage{psfrag}
\usepackage{graphicx}

\usepackage{amssymb, amsmath, amsbsy, bm, bbm}
	\allowdisplaybreaks[4]

	\newtheorem{Proposition}{Proposition}
	\newtheorem{Remark}{Remark}
	
\usepackage{lineno, hyperref}
	\hypersetup{hidelinks} 
\usepackage{algorithm, algpseudocode}
\usepackage{textcomp}
\usepackage[font={small}]{caption}
\usepackage[font={small}]{subcaption}
\usepackage{enumitem}
	\setenumerate[1]{itemindent=\parindent, itemsep=0pt, partopsep=0pt, parsep=\parskip, topsep=5pt}
\usepackage{color}
\usepackage{multirow}

\newcommand{\ppsi}{\boldsymbol{\psi}}
\newcommand{\PPsi}{\boldsymbol{\Psi}}

\DeclareMathOperator*{\argmax}{arg\,max\,}
\DeclareMathOperator*{\argmin}{arg\,min\,}
\DeclareMathOperator*{\cov}{cov}
\DeclareMathOperator*{\dd}{d\!}

\DeclareMathOperator*{\E}{E}

\DeclareMathOperator*{\var}{var}

\def\blue#1{\textcolor{blue}{#1}}

\begin{document}
\firstpage{1}
\lastpage{25}
\jvol{xx}
\issue{yy}
\jyear{2021}
\jid{CJS}
\aid{???}
\rhauthor{Zhou and Sang}
\copyrightline{Statistical Society of Canada}
\Frenchcopyrightline{Soci\'et\'e statistique du Canada}
\received{\rec{8}{May}{2020}}
\accepted{\acc{28}{Jan}{2021}}

\renewcommand{\eqref}[1]{(\ref{#1})}
\newcommand{\mb}[1]{\mathbf{#1}}
\newcommand{\mbb}[1]{\mathbb{#1}}
\newcommand{\mt}[1]{\mathrm{#1}}
\newcommand{\rv}{random variable}

\title[]{Continuum centroid classifier for functional data}
\author{Zhiyang Zhou\authorref{1}}
\author{Peijun Sang\authorref{2}\thanksref{*}}
\affiliation[1]{Department of Preventive Medicine, Northwestern University Feinberg School of Medicine, Chicago, IL 60611, United States}
\affiliation[2]{Department of Statistics \& Actuarial Science, University of Waterloo, ON N2L 3G1, Canada}

\correspondingauthor[*]{\\\email{peijun.sang@uwaterloo.ca}}

\startabstract{%
\begin{abstract}
\abstractsection{}
\abstractsection{Abstract}
 Aiming at the binary classification of functional data,
    we propose the continuum centroid classifier (CCC)
    built upon projections of functional data onto one specific direction.
    This direction is obtained via bridging the regression and classification.
    Controlling the extent of supervision,
    our technique is neither unsupervised nor fully supervised.
    Thanks to the intrinsic infinite dimension of functional data,
    one of two subtypes of CCC enjoys the (asymptotic) zero misclassification rate.
    Our proposal includes an effective algorithm
    that yields a consistent empirical counterpart of CCC.
    Simulation studies demonstrate the performance of CCC in different scenarios.
    Finally,
    we apply CCC to two real examples.
\abscopyright
\fabstractsection{}
\fabstractsection{R\'{e}sum\'{e}}
Ins\'{e}rer votre r\'{e}sum\'{e}
ici. We will supply a French abstract for those authors who
can't prepare it themselves.\Frenchabscopyright
\end{abstract}}
\makechaptertitle


\section{Introduction}\label{sec:introduction}

\setlength{\parskip}{0em}

With the development of technology,
data with complex structures
such as curves and images
have become more and more popular in various fields.
Functional data analysis (FDA)
handles one type of such data,
which take values over a continuum like space or time.
Typical examples of functional data include intraday yield curves for high frequency trading,
near-infrared spectra and blood signals measured in functional magnetic resonance imaging.
For a comprehensive overview of FDA,
one can refer to,
e.g. \cite{RamsaySilverman2005} and \cite{FerratyVieu2006}.

One important problem in FDA is the classification for functional data,
which is potentially applicable to,
for instance,
the diagnosis of multiple sclerosis (MS).
MS is an immune disorder which impairs the central nervous system.
Typical symptoms of MS include
fatigue, vision problems, sensation impairment and cognitive impairment.
Traditionally the diagnosis of MS relies on signs, symptoms and medical tests;
yet,
as reported by \cite{SbardellaTonaPetsasPantano2013},
some important results in MS study have been obtained from
applications of diffusion tensor imaging (DTI),
since DTI is a powerful tool in quantifying the demyelination and axonal loss
resulted from MS \citep{GoldsmithCrainiceanuCaffoReich2012}.
If DTI results are recorded for both healthy controls and MS patients,
the corresponding diagnosis can be interpreted as
classifying curves into two status,
i.e. a binary functional classification problem.
More details about this application will be presented in Section \ref{sec:numerical}.

In the literature of FDA,
there has been extensive research on functional data classification.
\cite{FerratyVieu2003} proposed a kernel estimator
of the posterior probability that a new curve belongs to a given class.
\cite{Shin2008} extended the linear discriminant analysis (LDA) to functional data.
The former work defines a distance between curves,
while the latter one makes use of the reproducing kernel Hilbert space
to define an inner product;
both of them are built upon the infinite dimensional space where functional data lie.
As pointed out by \cite{FanFengTong2012},
a large number of covariates would pose challenges for classification
in the multivariate context.
Functional data, however, are intrinsically infinite dimensional.
In light of this fact,
many researchers suggest reducing the dimension first
and then implementing classical classification algorithms in the reduced space.
Functional principal component (FPC) analysis is
one of the most commonly used approaches for dimension reduction.
Projections of functional data onto
the directions of functional principal components
are referred to as FPC scores in the literature.
\cite{GaleanoJosephLillo2015} proposed
both linear and quadratic Bayes classifiers on FPC scores.
\cite{DaiMullerYao2017} further extended this idea to
density ratios of FPC scores and
showed that the resulting classifier
is equivalent to the quadratic discriminant analysis (QDA) for Gaussian random curves.
Additionally,
a logistic regression on FPC scores was considered in \cite{LengMuller2005}
to discriminate  temporal gene regulation data.
\cite{rossi2006support} implemented the support vector machine on FPC scores.

Our work is mainly motivated by the centroid classifier proposed by
\cite{DelaigleHall2012}
who suggested projecting functional data to a specific direction first.
To identify this projection direction,
the authors converted the classification problem to
a functional linear regression problem.
The slope function in the functional linear model
is then taken as the projection direction they needed.
They proposed two methods in estimating the slope function:
one with FPC basis functions and
the other with the functional partial least squares (FPLS) basis functions.
The first method accounts for the variability of functional covariates
but ignores the information of the outcome when estimating the projection direction,
whereas the second one focuses on the covariance
between the outcome and the functional covariate.
As argued in \cite{Jung2018},
the projection direction from FPC basis
sometimes fails to capture the difference between mean functions,
while FPLS basis is prone to
sensitivity and vulnerability to small signals;
a combination of these two ideas
may be preferable than either of them.
We therefore propose a new method to estimate the projection direction,
which captures both variability of functional data
and covariance between the functional covariate and the outcome.
A distance-based centroid classifier is
then provided after projecting functional data onto this specific direction.
Moreover, under some regularity conditions,
we establish the asymptotic perfect classification property for
the newly proposed classifier.
Both simulation studies and real examples demonstrate that
our proposed classifier compares favorably with
the two methods given by \cite{DelaigleHall2012}
in terms of classification accuracy in finite samples.

The rest of this paper is organized as follows.
In \autoref{sec:ccc},
we introduce the two subtypes of our classifier,
including both the population and empirical versions,
and then establish the property of (asymptotically) perfect classification.
The numerical illustration in \autoref{sec:numerical}
investigates the performance of our proposal,
highlighting the settings in favor of it.
\autoref{sec:conclusion} gives concluding remarks as well as discussions.
Technical details are relegated to Appendix.

\section{Methodology and theory}
\label{sec:ccc}

\subsection{Formalization of the problem}
Suppose that $(X_1, Y_1),\ldots,(X_N, Y_N)$
are $n$ independently and identically distributed (iid)
copies of $(X, Y)$,
where $X$ is a random function defined on the interval
$\mathcal{T}=[t_{\min},t_{\max}]$,
and $Y$ is the label of $X$ taking values 0 or 1.
In other words,
each $X_i$ is sampled from
a mixture of two populations $\Pi_0$ and $\Pi_1$
with the indicator $Y_i=\mathbbm{1}(X_i \in \Pi_1)$.
Of interest is the binary classification for
a newly observed $X^*$ distributed as $X$ but independent of $X_1,\ldots,X_N$.
Assume that the sub-mean and sub-covariance functions for $\Pi_k$,
$k=0,1$,
are respectively
\begin{equation}\label{eq:mu.k}
	\mu_{[k]}(t)=\E\{X(t)\mid Y = k\}
\end{equation}
and, for all $s,t\in\mathcal{T}$,
\begin{equation}\label{eq:cov.fun.k}
    v_X^{[k]}(s,t)
    =\cov\{X(s),X(t)\mid Y = k\}.
\end{equation}
Let $\pi_0=\Pr(X_i\in\Pi_0)\in (0,1)$.
We have decomposition
$$
	\mu(\cdot)=\E X(\cdot)=\pi_0\mu_{[0]}(\cdot)+(1-\pi_0)\mu_{[1]}(\cdot)
$$
as well as,
according to the law of total covariance,
\begin{equation}\label{eq:cov.fun.total}
    v_X(s,t) = \cov\{X(s),X(t)\} = v_X^W(s,t)+v_X^B(s,t),
\end{equation}
where
\begin{align}
	v_X^W(s,t) &=\pi_0v_X^{[0]}(s,t)+(1-\pi_0)v_X^{[1]}(s,t), \label{eq:cov.within}
	\\
	v_X^B(s,t) &=\pi_0(1-\pi_0)\{\mu_{[1]}(s)-\mu_{[0]}(s)\}\{\mu_{[1]}(t)-\mu_{[0]}(t)\}\notag
\end{align}
are respectively the within- and between-group covariance functions.
Correspondingly,
define covariance operators
$\mathcal{V}_X, \mathcal{V}_X^{[k]}: L^2(\mathcal{T})\to L^2(\mathcal{T})$, $k=0,1$,
such that,
for $f\in L^2(\mathcal{T})$,
\begin{align*}
    \mathcal{V}_X(f)(\cdot) &=\int_{\mathcal{T}}f(s)v_X(s,\cdot)\dd s,
    \\
    \mathcal{V}_X^{[k]}(f)(\cdot) &=\int_{\mathcal{T}}f(s)v_X^{[k]}(s,\cdot)\dd s.
\end{align*}

Throughout this paper,
we abbreviate the Lebesgue integral $\int_{\mathcal{T}}f(t)\dd t$ to $\int_{\mathcal{T}} f$.
The functions involved in this paper
are limited to $L^2(\mathcal{T})$ (or $L^2(\mathcal{T}^2)$),
the collection of square integrable functions defined on $\mathcal{T}$ (or $\mathcal{T}^2$).
The square-integrability of $v_X$ at \eqref{eq:cov.fun.total}
(resp. $v_X^{[k]}$ at \eqref{eq:cov.fun.k})
implies that it possesses only a countable number of nonnegative eigenvalues
$\{\lambda_1,\lambda_2,\ldots\}$
(resp. $\{\lambda_{k,1},\lambda_{k,2},\ldots\}$),
with corresponding eigenfunctions
$\{\phi_1,\phi_2,\ldots\}$
(resp. $\{\phi_{k1},\phi_{k2},\ldots\}$),
$k=0,1$.
In addition,
$\|\cdot\|$ stands for the $L^2$-norm,
i.e.,
$\|f\|$ equals $(\int_{\mathcal{T}}f^2)^{1/2}$ for $f\in L^2(\mathcal{T})$
and $(\int_{\mathcal{T}}\int_{\mathcal{T}}f^2)^{1/2}$ if $f\in L^2(\mathcal{T}^2)$.

\subsection{Review of centroid classifier}\label{sec:cc}
Before proceeding to our proposal,
we first review the centroid classifier
proposed by \citet{DelaigleHall2012}.
They projected functional data onto the one-dimensional space
spanned by a given function $\omega\in L^2(\mathcal{T})$,
say ${\rm span}(\omega)$,
and then constructed the classifier with the projection.
Specifically,
they defined a classifier
\begin{multline}\label{eq:centroid.classifier}
	\mathcal{D}(X^*\mid\omega)
	= \left\{ \int_{\mathcal{T}}\frac{\omega}{\|\omega\|}(X^*-\mu_{[1]}) \right\}^2\\
		-\left\{ \int_{\mathcal{T}}\frac{\omega}{\|\omega\|}(X^*-\mu_{[0]}) \right\}^2
		+2\ln\frac{\pi_0}{1-\pi_0},
\end{multline}
where
$|\int_{\mathcal{T}}\omega(X^*-\mu_{[k]})|/\|\omega\|$,
the magnitude of the projection of $X^*-\mu_{[k]}$ onto ${\rm span}(\omega)$,
can be regarded as the distance from $X^*$ to $\mu_{[k]}$ \eqref{eq:mu.k}.
When $\mathcal{D}(X^*\mid\omega)$ is positive,
$X^*$ is thought to be closer to $\mu_{[0]}$ and hence assigned to $\Pi_0$ and vice versa.
Given $\omega$,
this principle is identical to LDA
assuming $\int_{\mathcal{T}}X\omega$
to be normally distributed conditional on $X\in\Pi_k$
with $\var(\int_{\mathcal{T}}X\omega\mid X\in\Pi_k)=\|\omega\|^2$ for each $k$,
viz. $\int_{\mathbb{T}}X\omega\mid X\in\Pi_k\sim
\mathcal{N}(\int_{\mathbb{T}}\mu_{[k]}\omega, \|\omega\|_2^2)$.

It remains to select a direction $\omega\in L^2(\mathcal{T})$
to optimize the misclassification rate
\begin{multline*}
	{\rm err}\{\mathcal{D}(X^*\mid\omega)\}
	    =\pi_0\Pr\{\mathcal{D}(X^*\mid\omega)<0\mid X^*\in\Pi_0\}
	    \\
	    +(1-\pi_0)\Pr\{\mathcal{D}(X^*\mid\omega)>0\mid X^*\in\Pi_1\}.
\end{multline*}
\cite{DelaigleHall2012} proposed taking $\omega=\beta_{p,{\rm FPC}}$ at \eqref{eq:beta.fpc}
(resp. $\beta_{p,{\rm FPLS}}$ at \eqref{eq:beta.fpls}),
corresponding to FPC 
(concentrated on $v_X^W$ at \eqref{eq:cov.within}) (resp. FPLS) basis,
with a positive integer $p$ tuned via cross-validation.
The resulting classifier
$\mathcal{D}(X^*\mid\beta_{p,{\rm FPC}})$
(resp. $\mathcal{D}(X^*\mid\beta_{p,{\rm FPLS}})$)
is abbreviated to be PCC (resp. PLCC).
More specifically, these two projection directions are defined as
\begin{equation}\label{eq:beta.fpc}
	\beta_{p,{\rm FPC}}=
		\argmin_{\beta\in{\rm span}(\phi_1^W,\ldots,\phi_p^W)}
		\E\left\{ Y-\E Y-\int_{\mathcal{T}}\beta(X-\E X) \right\}^2
\end{equation}
and
\begin{equation}\label{eq:beta.fpls}
	\beta_{p,{\rm FPLS}}=
		\argmin_{\beta\in{\rm span}(w_{1,{\rm FPLS}},\ldots,w_{p,{\rm FPLS}})}
		\E\left\{ Y-\E Y-\int_{\mathcal{T}}\beta(X-\E X) \right\}^2,
\end{equation}
where
$\phi_1^W,\ldots,\phi_p^W$ are the first $p$ eigenfunctions of
$v_X^W$ at \eqref{eq:cov.within}
and $w_{1,\text{PLS}},\ldots,w_{p,\text{PLS}}$
are solutions to the following sequential optimization problems:
given $w_{1,{\rm FPLS}},\ldots,w_{j-1,{\rm FPLS}}$,
define
$$
	w_{j,{\rm FPLS}}=\argmax_{w:\|w\|=1}{\cov}^2\left(Y,\int_{\mathcal{T}}Xw\right)
$$
subject to $\int_{\mathcal{T}}w\mathcal{V}_X(w_{l,{\rm FPLS}})=0$,
$l=1,\ldots,j-1$.
Minimizers \eqref{eq:beta.fpc} and \eqref{eq:beta.fpls},
yet restricted in different linear spaces,
are both slope functions of
the (constrained) optimal approximations to $Y$
by a linear functional of $X$.
Taking either of them as $\omega$ in \eqref{eq:centroid.classifier},
\cite{DelaigleHall2012} succeeded in bridging
the binary classification problem to the functional linear regression.

\subsection{Continuum centroid classifier}
As mentioned in \autoref{sec:introduction},
an intermediate state between FPC and FPLS bases may be preferred.
Fixing $\alpha\in[0,1)$,
\citet[][ Proposition 4]{Zhou2019} defined the functional continuum (FC) basis functions
as sequential constrained maximizers of
\begin{equation}\label{eq:T.j.alpha}
    T_{j,\alpha}(w)
    ={\cov}^2\left(Y^{[j,\alpha]}, \int_{\mathcal{T}}X^{[j,\alpha]}w\right)\cdot
    {\textstyle\var^{-1+\frac{\alpha}{1-\alpha}}}
        \left(\int_{\mathcal{T}}X^{[j,\alpha]}w\right),
\end{equation}
where
$$
    X^{[j,\alpha]}=
        X^{[j-1,\alpha]}
        -\textstyle{\var^{-1/2}}\left(
					\int_{\mathcal{T}}Xw_{j-1,\alpha}\right)
		\cdot
		\left(\int_{\mathcal{T}}X^{[1,\alpha]}w_{j-1,\alpha}\right)
		\cdot
		\mathcal{V}_X\left(w_{j-1,\alpha}\right),
$$
and, for $j=2, 3,\ldots$,
$$
	Y^{[j,\alpha]}=
	    Y^{[j-1,\alpha]}
	    -\E Y
	    -\int_{\mathcal{T}}X^{[1,\alpha]}\beta_{j-1,\alpha},
$$
with $X^{[1,\alpha]}=X-\E X$ and $Y^{[1,\alpha]}=Y-\E Y$.
Specifically,
given $w_{1,\alpha},\ldots,w_{j-1,\alpha}$,
the $j$th FC basis function is
\begin{equation}\label{eq:w.j}
	w_{j,\alpha}=\argmax_{w:\|w\|=1}T_{j,\alpha}(w).
\end{equation}
FC basis captures not only variation of $X$ but also the covariance between $X$ and $Y$;
thus this dimension reduction technique
lies midway between (unsupervised) FPC and (fully supervised) FPLS.
FC basis reduces to FPLS when $\alpha=1/2$
and becomes irrelevant to $Y$ as $\alpha$ diverges.
By tuning $\alpha$,
FC basis controls the extent of supervision
and is expected to be neither unsupervised nor too supervised.

The continuum centroid classifier (CCC) is defined by
substituting
\begin{equation}\label{eq:beta.p.alpha}
    \beta_{p,\alpha}=
		\argmin_{\beta\in{\rm span}(w_{1,\alpha}, \ldots,w_{p,\alpha})}
		\E\left\{ Y-\E Y-\int_{\mathcal{T}}\beta(X-\E X) \right\}^2
\end{equation}
for $\omega$ in \eqref{eq:centroid.classifier} and,
simultaneously,
dropping the assumption $\var(\int_{\mathcal{T}}X\omega\mid X\in\Pi_k)=\|\omega\|^2$.
In details,
CCC assigns a label to a random trajectory $X^*$
by respectively applying LDA and QDA
to $\int_{\mathcal{T}}X^*\beta_{p,\alpha}$
and hence owns two subtypes,
say CCC-L and CCC-Q:
CCC-Q is given by
\begin{multline}\label{eq:D.Q}
	\mathcal{D}_Q(X^*\mid\beta_{p,\alpha})
	=\sigma_{[1]}^{-2}(\beta_{p,\alpha})\left\{
	    \int_{\mathcal{T}}\beta_{p,\alpha}(X^*-\mu_{[1]})
	\right\}^2
	\\
	-\sigma_{[0]}^{-2}(\beta_{p,\alpha})\left\{
	    \int_{\mathcal{T}}\beta_{p,\alpha}(X^*-\mu_{[0]})
	\right\}^2
	+2\ln\frac{\pi_0\sigma_{[1]}(\beta_{p,\alpha})}{(1-\pi_0)\sigma_{[0]}(\beta_{p,\alpha})}
\end{multline}
with
\begin{equation}\label{eq:sigma.k}
    \sigma_{[k]}^2(\omega)=\var\left(\int_{\mathcal{T}}X\omega\mid X\in\Pi_k\right)
\end{equation}
for each $k$, and CCC-L by
\begin{multline}\label{eq:D.L}
	\mathcal{D}_L(X^*\mid\beta_{p,\alpha})
	=\left\{
	    \int_{\mathcal{T}}\beta_{p,\alpha}(X^*-\mu_{[1]})
	\right\}^2
	\\
	-\left\{
	    \int_{\mathcal{T}}\beta_{p,\alpha}(X^*-\mu_{[0]})
	\right\}^2
	+2\sigma^2(\beta_{p,\alpha})\ln\frac{\pi_0}{1-\pi_0}
\end{multline}
if one believes $\sigma^2(\omega)=\sigma_{[0]}^2(\omega)=\sigma_{[1]}^2(\omega)$.
Analogous to \eqref{eq:centroid.classifier},
positive $\mathcal{D}_L(X^*\mid\beta_{p,\alpha})$ (or $\mathcal{D}_Q(X^*\mid\beta_{p,\alpha})$)
suggests classifying $X^*$ to $\Pi_0$.


As long as \ref{cond:unbounded} in the appendix stands,
in theory one can expect an asymptotically perfect classification given by CCC-L;
see \autoref{prop:0.error}.
It is worth emphasizing that
\autoref{prop:0.error} does not require normality or
specific variance structure of the two sub-populations.

\begin{Proposition}\label{prop:0.error}
    Holding \ref{cond:unbounded},
    CCC-L asymptotically leads to no misclassification as $p\to\infty$.
\end{Proposition}

\subsubsection{Empirical implementation}\label{sec:implementation}

In general
it is impossible to observe entire trajectories.
In this sense,
the procedure of estimating $\beta_{p,\alpha}$ \eqref{eq:beta.p.alpha} in \citet{Zhou2019}
is not detailed enough:
the algorithm over there is not described in the matrix form.
We fix it in this section.

For brevity,
$X_i$'s are all assumed to be densely digitized on equispaced $M+1$ time points
$t_m=t_{\min}+(m - 1)\Delta t$, $m=1,\ldots,M+1$,
with $\Delta t = (t_{\max}-t_{\min})/M$.
Reformulating
the infinite-dimensional optimization problem \eqref{eq:w.j}
to a finite-dimensional one,
we employ the penalized cubic B-spline smoothing
(\citealp{RamsaySilverman2005}, Sections 5.2.4--5.2.5)
to each trajectory,
i.e.,
seeking for a surrogate of $X_i$ in
the $L$ \citep[$=M+3$, as recommended by][pp. 86]{RamsaySilverman2005}
dimensional linear space
\begin{equation}\label{eq:bs.k}
    BS_L={\rm span}(\psi_1,\ldots,\psi_L)
\end{equation}
spanned by cubic B-splines $\psi_1,\ldots,\psi_L$;
refer to, e.g., \citet[][ Chapter 4]{deBoor2001}, 
for more details on B-splines.
Specifically,
the estimator for the $i$th trajectory, $i=1,\ldots,N$, is
\begin{equation}\label{eq:x.i.hat}
    \hat{X}_i=\hat{\bm{c}}_i^\top\ppsi,
\end{equation}
where
\begin{align}
    \ppsi
    &=\ppsi(\cdot)=[\psi_1(\cdot),\ldots,\psi_L(\cdot)]^\top,
    \label{eq:ppsi}\\
    \hat{\bm{c}}_i
    &=(\PPsi^\top\PPsi+\theta_0\mathbf{Pen})^{-1}\PPsi^\top\bm{X}_i,
    \label{eq:c.i.hat}
\end{align}
with
\begin{align}
    \PPsi
    &=[\psi_k(t_m)]_{(M+1)\times L}=[\ppsi(t_1),\ldots,\ppsi(t_{M+1})]^\top,
    \label{eq:PPsi}\\
    \mathbf{Pen}
    &=\left[\int_{\mathcal{T}}\psi_{l_1}''\psi_{l_2}''\right]_{1\leq l_1, l_2\leq L},
    \label{eq:Pen}\\
    \bm{X}_i
    &=[X_i(t_1),\ldots,X_i(t_{M+1})]^\top.
    \notag
\end{align}
Positive smoothing parameter $\theta_0$
is communal for all $i$ and 
picked up automatically through the generalized cross-validation \citep[GCV,][]{CravenWahba1979},
i.e.,
$$
	\theta_0 = \argmin_{\theta}\frac{
		\sum_{i=1}^{N}\sum_{m=1}^{M+1}\{X_i(t_m)-\hat X_i(t_m)\}^2
	}{
		[M+1-{\rm trace}\{\PPsi(\PPsi^\top\PPsi+\theta\mathbf{Pen})^{-1}\PPsi^\top\}]^2.
	}
$$ 
Thanks to dense observations,
the presmoothing is able to recover underlying curves accurately (in the $L^2$ sense) under some regularity conditions.
Also, 
presmoothed curves are anticipated to enhance the classification accuracy \citep{DaiMullerYao2017}.

\begin{Proposition}\label{prop:x.i}
    Holding \ref{cond:x.derivative} and \ref{cond:eigenvalue} in the appendix,
    for each $i$,
    $\|\hat{X}_i-X_i\|\to 0$ in probability as $M\to\infty$.
\end{Proposition}

The empirical version of $T_{j,\alpha}$ \eqref{eq:T.j.alpha},
say $\hat{T}_{j,\alpha}$,
is then constructed by
replacing $X_i$, $\var(\cdot)$, and $\cov(\cdot)$ in \eqref{eq:T.j.alpha}
with respective empirical counterparts
or, identically,
substituting $\hat{X}_i$ \eqref{eq:x.i.hat} for $X_i$ in \citet[][ Proposition 5]{Zhou2019}.

\begin{Proposition}\label{prop:maximizer.T.hat}
    Suppose
    $\hat{w}_{j,\alpha}=\max_{w:\|w\|=1}\hat{T}_{j,\alpha}(w)$
    is an estimator of $w_{j,\alpha}$ \eqref{eq:w.j}.
    Then $\hat{w}_{j,\alpha}$ must lie in the linear space $BS_L$.
\end{Proposition}

Start with optimizing $\hat{T}_{1,\alpha}(w)$.
\autoref{prop:maximizer.T.hat} narrows down
our search from $\{w:\|w\|=1, w\in L^2(\mathcal{T})\}$ to
$\{w: w=\bm{b}^\top\ppsi, \bm{b}^\top\mathbf{W}\bm{b}=1, \bm{b}\in \mathbb{R}^{L\times 1}\}
=\{w:w=\bm{b}^\top\mathbf{W}^{-1/2}\ppsi, \bm{b}^\top\bm{b}=1, \bm{b}\in\mathbb{R}^{L\times 1}\}$
with invertible and symmetric
\begin{equation}\label{eq:W}
    \mathbf{W}
    =\left[\int_{\mathcal{T}}\psi_{l_1}\psi_{l_2}\right]_{1\leq l_1, l_2\leq L}.
\end{equation}
The maximization of $\hat{T}_{1,\alpha}(w)$
(subject to $\|w\|=1$)
is reformulated as $L$-dimensional optimization problem
\begin{equation}\label{eq:reform.optim}
    \max_{\bm{b}\in\mathbb{R}^{L\times1}}
    (\bm{b}^\top\mathbf{W}^{1/2}\hat{\mathbf{C}}_{\rm c}^\top\bm{Y}_{\rm c})^2
    (
        \bm{b}^\top\mathbf{W}^{1/2}\hat{\mathbf{C}}_{\rm c}^\top
	    \hat{\mathbf{C}}_{\rm c}\mathbf{W}^{1/2}\bm{b}
	)^{\alpha/(1-\alpha)-1}
\end{equation}
subject to $\bm{b}^\top\bm{b}=1$,
where,
with $\hat{\bm{c}}_i$ at \eqref{eq:c.i.hat},
\begin{align*}
    \hat{\mathbf{C}}_{\rm c}
    &=\left[\hat{\bm{c}}_1-\frac{1}{N}\sum_{i=1}^N\hat{\bm{c}}_i,
    \ldots,\hat{\bm{c}}_N-\frac{1}{N}\sum_{i=1}^N\hat{\bm{c}}_i\right]^\top,
    \\
    \bm{Y}_{\rm c}
    &=\left[Y_1-\frac{1}{N}\sum_{i=1}^N Y_i,\ldots,Y_N-\frac{1}{N}\sum_{i=1}^N Y_i\right]^\top.
\end{align*}
Note that
the solution to \eqref{eq:reform.optim}
is necessarily located in the row space of
$\hat{\mathbf{C}}_{\rm c}\mathbf{W}^{1/2}$,
i.e.,
the search region is further restricted to
$\{w:w=\bm{b}^\top\mathbf{V}^\top\mathbf{W}^{-1/2}\ppsi,
\bm{b}^\top\bm{b}=1, \bm{b}\in\mathbb{R}^{r\times 1}\}$,
where $r={\rm rank}(\hat{\mathbf{C}}_{\rm c}\mathbf{W}^{1/2})\leq\min\{L,N\}$
and $L\times r$ matrix $\mathbf{V}$ comes from
the thin singular value decomposition of
$\hat{\mathbf{C}}_{\rm c}\mathbf{W}^{1/2}$:
$\hat{\mathbf{C}}_{\rm c}\mathbf{W}^{1/2}$
is decomposed into
$\mathbf{U}\mathbf{R}\mathbf{V}^\top$
with an invertible diagonal matrix $\mathbf{R}$
and semi-orthogonal matrices $\mathbf{U}$ and $\mathbf{V}$
such that $\mathbf{U}^\top\mathbf{U}=\mathbf{V}^\top\mathbf{V}=\mathbf{I}_r$.
In this way the dimension of \eqref{eq:reform.optim} is further reduced to $r$.

Write $\mathbf{G}_1=\mathbf{U}\mathbf{R}$.
The estimator of the first FC basis function then takes the form
$$
    \hat{w}_{1,\alpha}
    =\bm{b}_{1,\alpha}^\top\mathbf{V}^\top\mathbf{W}^{-1/2}\ppsi
$$
in which
$$
    \bm{b}_{1,\alpha}
    =\argmax_{\bm{b}\in\mathbb{R}^{r\times 1}:\bm{b}^\top\bm{b}=1}
        (\bm{b}^\top\mathbf{G}_1^\top\bm{Y}_{\rm c})^2
        (
            \bm{b}^\top\mathbf{G}_1^\top\mathbf{G}_1\bm{b}
        )^{\alpha/(1-\alpha)-1}.
$$
Subsequently and successively,
for $j\geq 2$,
given $j - 1$ vectors $\bm{b}_{1,\alpha},\ldots,\bm{b}_{j-1,\alpha}$,
we just have to replace previous $\mathbf{G}_1$
with deflated $\mathbf{G}_j = \mathbf{P}_{j-1}\mathbf{G}_1$,
where
$\mathbf{P}_0=\mathbf{I}_N$ and
$
    \mathbf{P}_{j-1}
    =\mathbf{I}_N-\mathbf{H}_{j-1}
        (\mathbf{H}_{j-1}^\top\mathbf{H}_{j-1})^{-1}
        \mathbf{H}_{j-1}^\top
$
is the projection matrix associated with
the orthogonal complement of column space of
$$
    \mathbf{H}_{j-1}
    =\hat{\mathbf{C}}_{\rm c}\mathbf{W}^{1/2}
        [\mathbf{V}\bm{b}_{1,\alpha},\ldots,\mathbf{V}\bm{b}_{j-1,\alpha}]
    =\mathbf{U}\mathbf{R}[\bm{b}_{1,\alpha},\ldots,\bm{b}_{j-1,\alpha}].
$$
Namely,
for all $j$,
\begin{equation}\label{eq:w.j.hat}
    \hat{w}_{j,\alpha}
    =\bm{b}_{j,\alpha}^\top\mathbf{V}^\top\mathbf{W}^{-1/2}\ppsi,
\end{equation}
with
\begin{align}
    \bm{b}_{j,\alpha}
    &=\argmax_{\bm{b}\in\mathbb{R}^{r\times 1}:\bm{b}^\top\bm{b}=1}
        (\bm{b}^\top\mathbf{G}_j^\top\bm{Y}_{\rm c})^2
        (
            \bm{b}^\top\mathbf{G}_j^\top\mathbf{G}_j\bm{b}
        )^{\alpha/(1-\alpha)-1}
    \label{eq:b.j}\\
    &=\left\{
        \bm{Y}_{\rm c}^\top\mathbf{G}_j
        (
            \mathbf{G}_j^\top\mathbf{G}_j
            +\delta_{j,\alpha}^{-1}\zeta_{j,\alpha}\mathbf{I}_r
        )^{-2}
        \mathbf{G}_j^\top\bm{Y}_{\rm c}
    \right\}^{-1/2}
    (
        \mathbf{G}_j^\top\mathbf{G}_j
        +\delta_{j,\alpha}^{-1}\zeta_{j,\alpha}\mathbf{I}_r
    )^{-1}
    \mathbf{G}_j^\top\bm{Y}_{\rm c}
    \label{eq:b.j.ridge}
\end{align}
in which
$\zeta_{j,\alpha}$ is the largest eigenvalue of $\mathbf{G}_j^\top\mathbf{G}_j$.
The ridge-type \eqref{eq:b.j.ridge} is deduced from \citet[][ Proposition 2.1]{BjorkstromSundberg1999}.
The only unknown $\delta_{j,\alpha}$ in \eqref{eq:b.j.ridge}
is the local maximizer in $(-1,\infty)\setminus\{0\}$ of the univariate function
\begin{align*}
	Q_{j,\alpha}(\delta)=
	&\left\{
		\bm{Y}_{\rm c}^\top\mathbf{G}_j
        (
            \mathbf{G}_j^\top\mathbf{G}_j
            +\delta^{-1}\zeta_{j,\alpha}\mathbf{I}_r
        )^{-1}
        \mathbf{G}_j^\top\bm{Y}_{\rm c}
    \right\}^2
	\\
	&\times
	\left\{
		\bm{Y}_{\rm c}^\top\mathbf{G}_j
        (
            \mathbf{G}_j^\top\mathbf{G}_j
            +\delta^{-1}\zeta_{j,\alpha}\mathbf{I}_r
        )^{-2}
        \mathbf{G}_j^\top\bm{Y}_{\rm c}
	\right\}^{\alpha/(\alpha-1)}
	\\
	&\times
    \big\{
		\bm{Y}_{\rm c}^\top\mathbf{G}_j
        (
            \mathbf{G}_j^\top\mathbf{G}_j
            +\delta^{-1}\zeta_{j,\alpha}\mathbf{I}_r
        )^{-1}
        \mathbf{G}_j^\top\mathbf{G}_j
    \\
    &\qquad
        (
            \mathbf{G}_j^\top\mathbf{G}_j
            +\delta^{-1}\zeta_{j,\alpha}\mathbf{I}_r
        )^{-1}
        \mathbf{G}_j^\top\bm{Y}_{\rm c}
    \big\}^{\alpha/(1-\alpha)-1}
\end{align*}
and expected to be figured out through an arbitrary computer algebra system,
where
$Q_{j,\alpha}(\delta)$ is obtained by
plugging \eqref{eq:b.j.ridge} back into the maximization objective in \eqref{eq:b.j}.
Fixing $p$,
we then proceed to an estimator for $\beta_{p,\alpha}$ \eqref{eq:beta.p.alpha}:
\begin{align*}
    \hat{\beta}_{p,\alpha}
    &=[\hat{w}_{1,\alpha},\ldots,\hat{w}_{p,\alpha}]
		(\mathbf{H}_p^\top\mathbf{H}_p)^{-1}\mathbf{H}_p^\top\bm{Y}_{\rm c}
	\\
	&=\ppsi^\top\mathbf{W}^{-1/2}\mathbf{V}
	    [\bm{b}_{1,\alpha},\ldots,\bm{b}_{p,\alpha}]
	    (\mathbf{H}_p^\top\mathbf{H}_p)^{-1}\mathbf{H}_p^\top\bm{Y}_{\rm c}.
\end{align*}

\begin{Remark}
   Despite the possible ambiguity in representing $\bm{b}_{j,\alpha}$ \eqref{eq:b.j.ridge},
   the consistence of $\hat{w}_{j,\alpha}$ \eqref{eq:w.j.hat}
   is not affected,
   as long as \ref{cond:unique} in the appendix is fulfilled;
   refer to \citet[][ Remark 1]{Zhou2019}.
\end{Remark}

Analogous to $X_i$'s,
the trajectory to be assigned, $X^*$,
is discretely observed and has to be estimated by
$$
    \hat{X}^*=\hat{\bm{c}}^{*\top}\ppsi,
$$
where $\hat{\bm{c}}^*$ is available by
applying the B-spline smoothing to $X^*(t_1),\ldots,X^*(t_M)$.
Let $N_0$ (resp. $N_1$)
denote the number of training trajectories belonging to $\Pi_0$ (resp. $\Pi_1$).
Estimating mean functions $\mu_{[k]}$ by
\begin{equation}\label{eq:mu.k.hat}
    \hat{\mu}_{[k]}
    =\frac{1}{N_k}\sum_{i=1}^N\hat{X}_i\mathbbm{1}(X_i\in\Pi_k)
    =\frac{1}{N_k}\sum_{i=1}^N\hat{\bm{c}}_i^\top\ppsi\mathbbm{1}(X_i\in\Pi_k),
\end{equation}
the empirical CCC-Q and -L are then given by,
respectively,
\begin{align}
	\hat{\mathcal{D}}_Q(\hat{X}^*\mid\hat{\beta}_{p,\alpha})
	=&\ \hat{\sigma}_{[1]}^{-2}(\hat{\beta}_{p,\alpha})\left\{
	    \int_{\mathcal{T}}\hat{\beta}_{p,\alpha}(\hat{X}^*-\hat{\mu}_{[1]})
	\right\}^2
	\notag
	\\
	&-\hat{\sigma}_{[0]}^{-2}(\hat{\beta}_{p,\alpha})\left\{
	    \int_{\mathcal{T}}\hat{\beta}_{p,\alpha}(\hat{X}^*-\hat{\mu}_{[0]})
	\right\}^2
	+2\ln\frac{N_0\hat{\sigma}_{[1]}(\hat{\beta}_{p,\alpha})}
	    {N_1\hat{\sigma}_{[0]}(\hat{\beta}_{p,\alpha})}
	\label{eq:D.Q.hat}
	\\
	\intertext{and}
	\hat{\mathcal{D}}_L(\hat{X}^*\mid\hat{\beta}_{p,\alpha})
	=&\ \hat{\sigma}_{\rm pool}^{-2}(\hat{\beta}_{p,\alpha})\left\{
	    \int_{\mathcal{T}}\hat{\beta}_{p,\alpha}(\hat{X}^*-\hat{\mu}_{[1]})
	\right\}^2
	\notag
	\\
	&-\hat{\sigma}_{\rm pool}^{-2}(\hat{\beta}_{p,\alpha})\left\{
	    \int_{\mathcal{T}}\hat{\beta}_{p,\alpha}(\hat{X}^*-\hat{\mu}_{[0]})
	\right\}^2
	+2\ln\frac{N_0}{N_1},
	\label{eq:D.L.hat}
\end{align}
where, for $k=0,1$,
\begin{align}
    \hat{\sigma}_{\rm pool}^2(\omega)
    &=(N-2)^{-1}\sum_{k=0}^1\sum_{i=1}^N\left\{
    \int_{\mathcal{T}}\omega(\hat{X}_i-\hat{\mu}_{[k]})
    \right\}^2\mathbbm{1}(X_i\in\Pi_k),
    \label{eq:sigma.pool.hat}\\
    \hat{\sigma}_{[k]}^2(\omega)
    &=(N_k-1)^{-1}\sum_{i=1}^N\left\{
    \int_{\mathcal{T}}\omega(\hat{X}_i-\hat{\mu}_{[k]})
    \right\}^2\mathbbm{1}(X_i\in\Pi_k).
    \label{eq:sigma.k.hat}
\end{align}

\begin{Proposition}\label{prop:ccc.converge}
    Fix $p\in\mathbb{Z}^+$ and $\alpha\in[0,1)$
    and assume \ref{cond:x.derivative}--\ref{cond:unique}.
    Empirical classifier $\hat{\mathcal{D}}_Q(\hat{X}^*\mid\hat{\beta}_{p,\alpha})$ \eqref{eq:D.Q.hat}
    (resp. $\hat{\mathcal{D}}_L(\hat{X}^*\mid\hat{\beta}_{p,\alpha})$ \eqref{eq:D.L.hat})
    converges to its population version
    $\mathcal{D}_Q(X^*\mid\beta_{p,\alpha})$ \eqref{eq:D.Q}
    (resp. $\mathcal{D}_L(X^*\mid\beta_{p,\alpha})$ \eqref{eq:D.L})
    in probability as $N$ diverges.
    Further,
    if \ref{cond:unbounded} holds too,
    a direct corollary is then
    $$
        \lim_{p\to\infty}\lim_{N\to\infty}
        {\rm err}\{\hat{\mathcal{D}}_L(\hat{X}^*\mid\hat{\beta}_{p,\alpha})\}
        =0.
    $$
\end{Proposition}

\subsubsection{Tuning parameter selection}\label{sec:tuning}
As explained in Section \ref{sec:cc},
this functional classification problem is convertible to a functional linear regression problem
when estimating the projection direction.
Alternative to the cross-validation,
GCV is frequently used
in choosing tuning parameters in functional linear models
\citep[e.g.,][]{CardotFerratySarda2003,
    CardotCrambesKneipSarda2007, ReissOgden2010}.
This GCV-based method is considerably more efficient than cross-validation in computation.
In this paper,
we suggest searching for the optimal pair of $(p,\alpha)$ by
respectively minimizing
\begin{align*}
    {\rm GCV}(p,\alpha)
        &=\frac{\sum_{i=1}^N\left[
            Y_i-\mathbbm{1}\{\hat{\mathcal{D}}_Q(\hat{X}_i\mid\hat{\beta}_{p,\alpha})< 0\}
        \right]^2}{(N-p-2)^2}
    \\
    \intertext{and}
    {\rm GCV}(p,\alpha)
        &=\frac{\sum_{i=1}^N\left[
            Y_i-\mathbbm{1}\{\hat{\mathcal{D}}_L(\hat{X}_i\mid\hat{\beta}_{p,\alpha})< 0\}
        \right]^2}{(N-p-2)^2}
\end{align*}
for CCC-Q and -L with respect to $(p,\alpha)$,
where the digit $2$ in parenthesis in the denominator corresponds to the number of populations.
\autoref{alg} details the implementation of CCC with the proposed GCV-based tuning scheme.

Rather than the usual rectanglar search grid,
the candidate pool for $(p,\alpha)$ here,
say $\{(p,\alpha): p\in\{1,\ldots, p_{\max,\alpha}\}, \alpha\in\{\alpha_1,\ldots,\alpha_J\}\subset [0,1)\}$,
is nonrectanglar and set up in a random way:
for each $\alpha\in\{\alpha_1,\ldots,\alpha_J\}$
($=\{0\times 10^{-1},\ldots,9\times 10^{-1},1-10^{-2},\ldots,1-10^{-4}\}$
in \autoref{sec:numerical}),
$p_{\max,\alpha}$
is randomly picked up from ${\rm Unif}\{1,\ldots,p_{\rm upper}\}$,
where $p_{\rm upper}$ can be determined by
\begin{equation}\label{eq:p.upper}
	p_{\rm upper}
	=\min\left\{j_0\in\mathbb{Z}^+:
	    \sum_{j=1}^{j_0}\hat{\lambda}_j^W/\sum_{j=1}^{\infty}\hat{\lambda}_j^W\geq 99\%\right\}
\end{equation}
in which $\hat{\lambda}_j^W$ estimates the $j$th top eigenvalue of
$v_X^W$ at \eqref{eq:cov.within}.
The random grid is typically of smaller cardinality and
hence leads to less running time.
\citet[][ Section 5]{Zhou2019} illustrated that
this strategy was accompanied with little loss in prediction accuracy.

\begin{algorithm}[t!]
	\caption{CCC tuned via GCV}
	\label{alg}
	\begin{algorithmic}[1]
		\State $p_{\max}\gets$ upper bound of number of FC basis functions.
		\For {$\alpha$ in a finite set}
		    \For {$p$ from 1 to $p_{\max,\alpha}$}
    		        \If {$p=1$}
    					\State $\mathbf{P}\gets\mathbf{I}_N$
    					\State $\mathbf{U}\mathbf{R}\mathbf{V}^\top\gets$ thin singular value decomposition of $\hat{\mathbf{C}}_{\rm c}\mathbf{W}^{1/2}$
            	        \State $\mathbf{G}_1\gets\mathbf{U}\mathbf{R}$
    				\Else
    					\State $\mathbf{P}\gets
    						\mathbf{P} \{\mathbf{I}_N-
    						\mathbf{G}_{p-1}\bm{b}_{p-1,\alpha}
    						(\bm{b}_{p-1,\alpha}^\top
    						    \mathbf{G}_{p-1}^\top\mathbf{G}_{p-1}
    						    \bm{b}_{p-1,\alpha}
    						)^{-1}
    						\bm{b}_{p-1,\alpha}^\top\mathbf{G}_{p-1}^\top\}$
    				\EndIf
            	    \State $\mathbf{G}_p\gets\mathbf{P}\mathbf{G}_1$
    				\State $\zeta\gets$ largest eigenvalue of
    				    $\mathbf{G}_p^\top\mathbf{G}_p$
    			    \State $\mathbf{L}(\delta)\gets
    			        (\mathbf{G}_p^\top\mathbf{G}_p+\delta^{-1}\zeta\mathbf{I}_r)^{-1}$
                    \State $Q(\delta)\gets\frac{
	                        \{
	        			        \bm{Y}_{\rm c}^\top\mathbf{G}_p\mathbf{L}(\delta)
	        			        \mathbf{G}_p^\top\bm{Y}_{\rm c}
	    			        \}^2
	                        \{
	        			        \bm{Y}_{\rm c}^\top\mathbf{G}_p\mathbf{L}^2(\delta)
	        			        \mathbf{G}_p^\top\bm{Y}_{\rm c}
	    			        \}^{\alpha/(1-\alpha)}
                    	}{
                    		\{
                    	        			        \bm{Y}_{\rm c}^\top\mathbf{G}_p\mathbf{L}(\delta)
                    	        			        \mathbf{G}_p^\top\mathbf{G}_p\mathbf{L}(\delta)
                    	        			        \mathbf{G}_p^\top\bm{Y}_{\rm c}
                    	    			        \}^{1-\{\alpha/(1-\alpha)\}}
                    	}$
    			    \State $\delta_{p,\alpha}\gets
    				    \argmin_{\delta\in(-1,0)\cup(0,\infty)}-\ln Q(\delta)$
    				\State $\bm{b}_{p,\alpha}\gets
    				        \mathbf{L}(\delta_{p,\alpha})\mathbf{G}_p^\top\bm{Y}_{\rm c}/
    				        \{
    				            \bm{Y}_{\rm c}^\top\mathbf{G}_p
    				            \mathbf{L}^2(\delta_{p,\alpha})\mathbf{G}_p^\top\bm{Y}_{\rm c}
    				        \}^{1/2}
    				    $
    				\State $\hat{w}_{p,\alpha}\gets
    				    \bm{b}_{p,\alpha}^\top\mathbf{V}^\top\mathbf{W}^{-1/2}\ppsi$
    		        \If {$p=1$}
        				\State $\hat{\beta}_{p,\alpha}\gets
        				    N^{-1/2}
                                (
                                    \bm{b}_{p,\alpha}^\top\mathbf{G}_p^\top\bm{Y}_{\rm c}
                                )
                                (
                                    \bm{b}_{p,\alpha}^\top\mathbf{G}_p^\top\mathbf{G}_p\bm{b}_{p,\alpha}
                                )^{-1/2}
                                \hat{w}_{p,\alpha}
        				$
    				\Else
        				\State $\hat{\beta}_{p,\alpha}\gets\hat{\beta}_{p-1,\alpha}+
        				    N^{-1/2}
                                (
                                    \bm{b}_{p,\alpha}^\top\mathbf{G}_p^\top\bm{Y}_{\rm c}
                                )
                                (
                                    \bm{b}_{p,\alpha}^\top\mathbf{G}_p^\top\mathbf{G}_p\bm{b}_{p,\alpha}
                                )^{-1/2}
                                \hat{w}_{p,\alpha}
        				$
    				\EndIf
	\algstore{alg}
	\end{algorithmic}
\end{algorithm}

\begin{algorithm}[t!]
	\begin{algorithmic}[1]
	\algrestore{alg}
    			\For {$i$ from 1 to $N$}
    			    \State $\int_{\mathcal{T}}\hat{\beta}_{p,\alpha}\hat{X}_i\gets
    			            N^{-1/2}\left\{
                                    \sum_{j=1}^p
                                    (
                                        \bm{b}_{j,\alpha}^\top\mathbf{G}_j\bm{Y}_{\rm c}
                                    )
                                    (
                                        \bm{b}_{j,\alpha}^\top\mathbf{G}_j^\top\mathbf{G}_j\bm{b}_{j,\alpha}
                                    )^{-1/2}
                                    \bm{b}_{j,\alpha}^\top\mathbf{V}_j^\top
                                \right\}
                                \mathbf{W}^{1/2}\hat{c}_i
    			        $
    			 \EndFor
    			 \State $\int_{\mathcal{T}}\hat{\beta}_{p,\alpha}\hat{\mu}_{[k]}\gets
    			        {\rm mean}\{
    			            \int_{\mathcal{T}}\hat{\beta}_{p,\alpha}\hat{X}_i\mathbbm{1}(X_i\in\Pi_k):i=1,\ldots,N
    			        \}
    			    $
    			 \State $\hat{\sigma}_{[k]}^2(\hat{\beta}_{p,\alpha})\gets
    			        \var\{
    			            \int_{\mathcal{T}}\hat{\beta}_{p,\alpha}\hat{X}_i\mathbbm{1}(X_i\in\Pi_k):i=1,\ldots,N
    			        \}
    			    $
    			 \State $\hat{\sigma}_{\rm pool}^2(\hat{\beta}_{p,\alpha})\gets
    			        (N-2)^{-1}
    			        \{(N_0-1)\hat{\sigma}_{[0]}^2(\hat{\beta}_{p,\alpha})
    			        +(N_1-1)\hat{\sigma}_{[1]}^2(\hat{\beta}_{p,\alpha})\}$
    			 \For {$i$ from 1 to $N$}
    			    \State $\hat{\mathcal{D}}_Q(\hat{X}_i\mid\hat{\beta}_{p,\alpha})\gets
    			            \hat{\sigma}_{[1]}^{-2}(\hat{\beta}_{p,\alpha})
    			                (\int_{\mathcal{T}}\hat{\beta}_{p,\alpha}\hat{X}_i
    			                -\int_{\mathcal{T}}\hat{\beta}_{p,\alpha}\hat{\mu}_{[1]})^2$
    			    \State\qquad $-\hat{\sigma}_{[0]}^{-2}(\hat{\beta}_{p,\alpha})
    			                (\int_{\mathcal{T}}\hat{\beta}_{p,\alpha}\hat{X}_i
    			                -\int_{\mathcal{T}}\hat{\beta}_{p,\alpha}\hat{\mu}_{[0]})^2
    			    			+2\ln\frac{N_0\hat{\sigma}_{[1]}(\hat{\beta}_{p,\alpha})}
	                            {N_1\hat{\sigma}_{[0]}(\hat{\beta}_{p,\alpha})}$
    			    \State $\hat{\mathcal{D}}_L(\hat{X}_i\mid\hat{\beta}_{p,\alpha})\gets
    			            \hat{\sigma}_{\rm pool}^{-2}(\hat{\beta}_{p,\alpha})
    			                (\int_{\mathcal{T}}\hat{\beta}_{p,\alpha}\hat{X}_i
    			                -\int_{\mathcal{T}}\hat{\beta}_{p,\alpha}\hat{\mu}_{[1]})^2$
    			    \State\qquad $-\hat{\sigma}_{\rm pool}^{-2}(\hat{\beta}_{p,\alpha})
    			                (\int_{\mathcal{T}}\hat{\beta}_{p,\alpha}\hat{X}_i
    			                -\int_{\mathcal{T}}\hat{\beta}_{p,\alpha}\hat{\mu}_{[0]})^2
    			            +2\ln(N_0/N_1)$
                \EndFor
			\EndFor
			\State ${\rm GCV}(p,\alpha)\leftarrow
					\sum_{i=1}^N
					    [Y_i-\mathbbm{1}\{
					    	\hat{\mathcal{D}}_Q(\hat{X}_i\mid\hat{\beta}_{p,\alpha})< 0
	                    \}]^2/(N-p-2)^2$
            \State\qquad or $\sum_{i=1}^N
					    [Y_i-\mathbbm{1}\{
	                    	\hat{\mathcal{D}}_L(\hat{X}_i\mid\hat{\beta}_{p,\alpha})< 0
	                    \}]^2/(N-p-2)^2
                $
		\EndFor
		\State $(p_{\rm opt},\alpha_{\rm opt})\leftarrow\argmin_{(p,\alpha)}{\rm GCV}(p,\alpha)$
	\end{algorithmic}
\end{algorithm}

\section{Numerical illustration}\label{sec:numerical}

In spite of theoretical arguments illustrating
the asymptotically perfect classification of CCC-L
in specific cases,
we were still in need of more evidences to support our proposals,
especially CCC-Q.
We then resorted to numerical studies
in order to compare the performance of PCC, PLCC and two CCC classifiers
in finite-sample applications.
Numbers of FPC (resp. FPLS) basis functions for PCC (resp. PLCC)
was selected from $\{1,\ldots,p_{\rm upper}\}$
through 5-fold cross-validation.
Two more classifiers
(viz. functional versions of logit regression and naive bayes)
were involved too in the comparison
and both implemented through
R-package \texttt{fda.usc} \citep{R-fda.usc}.
Our code trunks are publicly available at 
\url{https://github.com/ZhiyangGeeZhou/CCC}.

\subsection{Simulation study}\label{sec:simulation}

We generated $R=200$ samples,
each containing $N=200$ curves $X_i$, $i=1,\ldots, 200$.
In each sample,
we randomly preserved $80\%$ curves for training
and left for testing.
Each curve was spotted at 101 equally spaced points in $\mathcal{T}=[0,1]$,
i.e., $\{0, 1/100,\ldots,99/100,1\}$,
and iid as
$$
    X_i = \sum_{k=0}^1
        \left(
            \sum_{j=1}^5\lambda_{k,j}^{1/2}Z_{ij}\phi_{k,j} + \mu_{[k]}
        \right)
        \mathbbm{1}(X_i\in\Pi_k).
$$
Without loss of generality,
the difference of two mean functions
was set to be exactly $\mu_{[1]}$,
i.e., $\mu_{[0]}(\cdot)\equiv 0$.
Instead of a mixture of Gaussian processes,
a setup without normality was considered:
$Z_{ij}\sim\exp(1)-1$, $i=1,\ldots,N$, $j=1,\ldots,5$.
Although sub-covariance functions $v_X^{[0]}$ and $v_X^{[1]}$
shared the identical nonzero eigenvalues
200, 100, 1, .2, .1,
they might differ in eigenfunctions;
specifically,
we took the (normed-to-one) $j$th-order shifted Legendre polynomial
\citep[see, e.g.,][pp.~773--774]{Hochstrasser1972}
as the $j$th eigenfunction of $v_X^{[0]}$,
i.e.,
\begin{align*}
    \phi_{0,1}(t)
    &= \sqrt{3}(2t-1),
    \\
    \phi_{0,2}(t)
    &= \sqrt{5}(6t^2-6t+1),
    \\
    \phi_{0,3}(t)
    &= \sqrt{7}(20t^3-30t^2+12t-1),
    \\
    \phi_{0,4}(t)
    &= 3(70t^4-140t^3+90t^2-20t+1),
    \\
    \phi_{0,5}(t)
    &= \sqrt{11}(252t^5-630t^4+560t^3-210t^2+30t-1).
\end{align*}
As a result,
$p_{\rm upper}$,
the upper bound of number of components,
was set to 5 directly
(rather than following \eqref{eq:p.upper}).
We meanwhile accounted for two sorts of combinations of
$\mu_{[1]}$ and $\phi_{1,j}$:
\begin{enumerate}[label=(\roman*)]
	\item
	    $\mu_{[1]} = \rho\lambda_{1,1}^{1/2}\phi_{1,1}$
	    and $\phi_{0,j}=\phi_{1,j}$,
	    \label{case:simu1}
	\item
	    $\mu_{[1]} = \rho\lambda_{1,3}^{1/2}\phi_{1,3}$
	    and $\phi_{0,5-j}=\phi_{1,j}$.
	    \label{case:simu2}
\end{enumerate}
In both scenarios,
$\rho$ ($=1,10$) controlled
the magnitude of gap $\mu_{[1]}-\mu_{[0]}$
as well as 
the ratio of
between-group variation
to within-group one.
Despite the common adoption of $\pi_0=\Pr(X_i\in\Pi_0)=50\%$ in literature,
we checked as well unbalanced arrangement like $\pi_0=80\%$.
These in total eight setting combinations
would help us clarify the joint impact from
$\rho$, $\pi_0$ and $\mu_{[1]}-\mu_{[0]}$
on misclassification.

\begin{figure}[t!]
	\centering
	\begin{subfigure}{.45\textwidth}
		\centering
		\includegraphics[width=\textwidth, height=.3\textheight]
		    {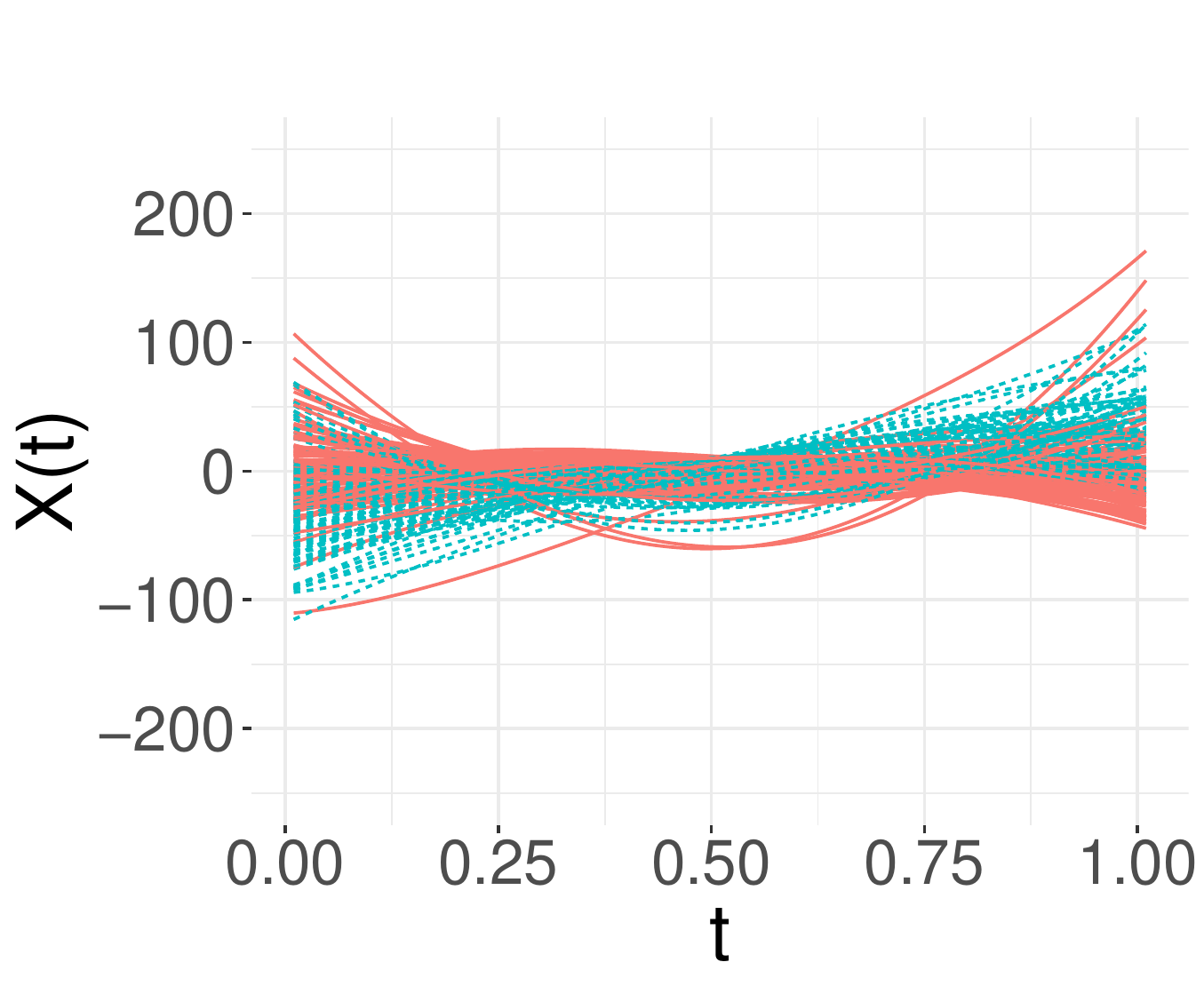}
		\caption{\centering Design \ref{case:simu1} with $\rho=1$}
		\label{fig:sample.design1&rho1}
	\end{subfigure}
	\begin{subfigure}{.45\textwidth}
		\centering
		\includegraphics[width=\textwidth, height=.3\textheight]
		    {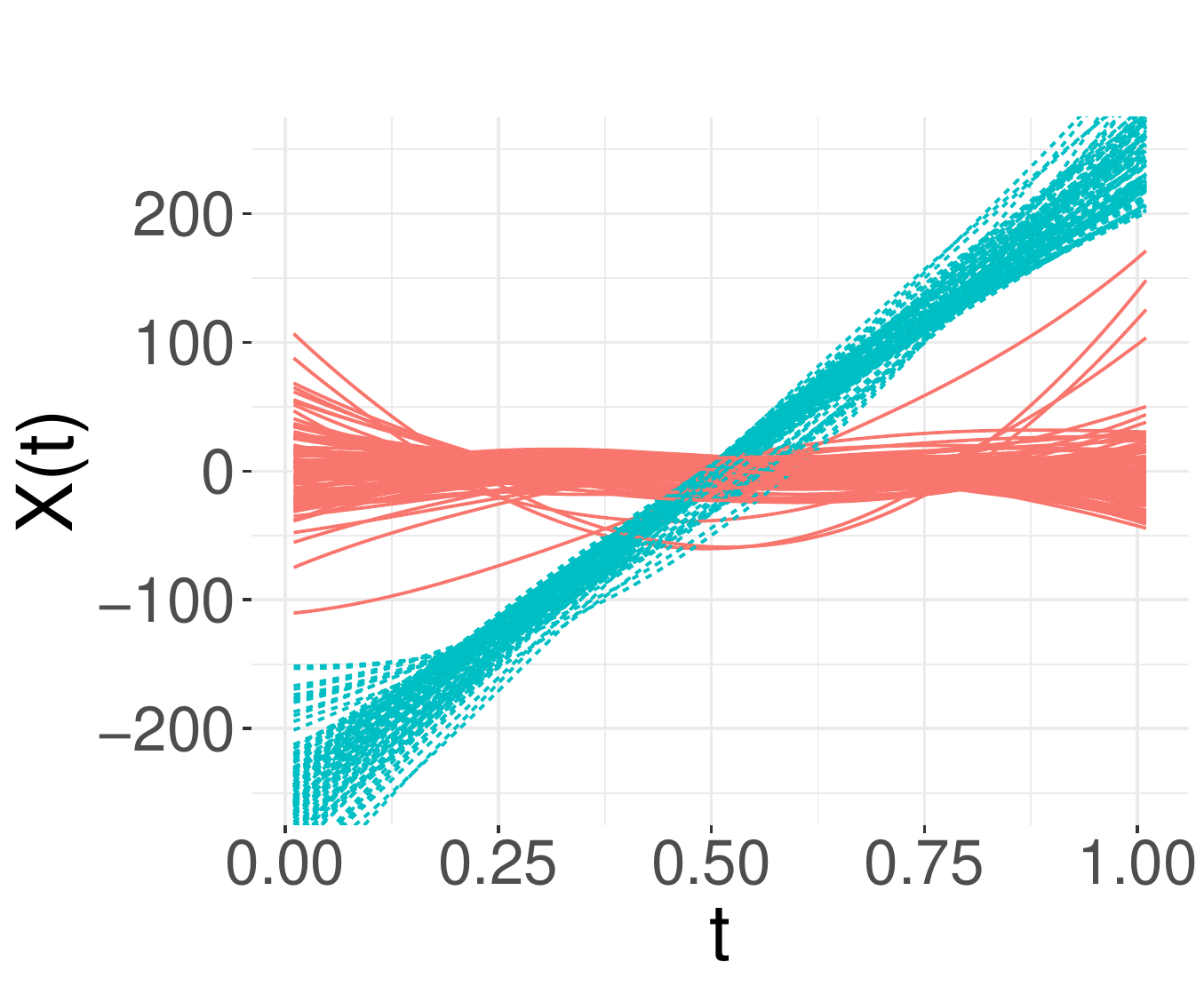}
		\caption{\centering Design \ref{case:simu1} with $\rho=10$}
	\end{subfigure}
	
	\begin{subfigure}{.45\textwidth}
		\centering
		\includegraphics[width=\textwidth, height=.3\textheight]
		    {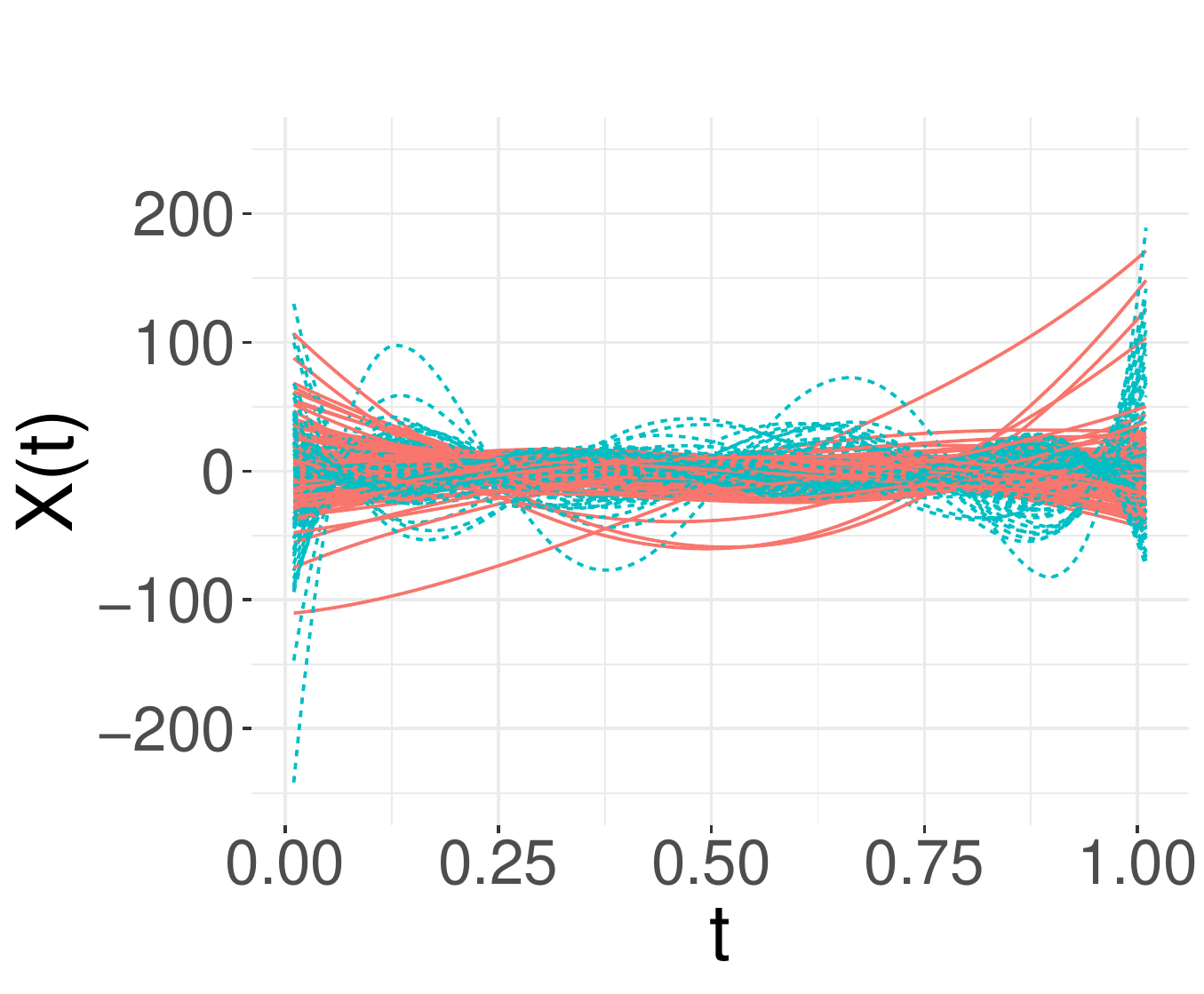}
		\caption{\centering Design \ref{case:simu2} with $\rho=1$}
	\end{subfigure}
	\begin{subfigure}{.45\textwidth}
		\centering
		\includegraphics[width=\textwidth, height=.3\textheight]
		    {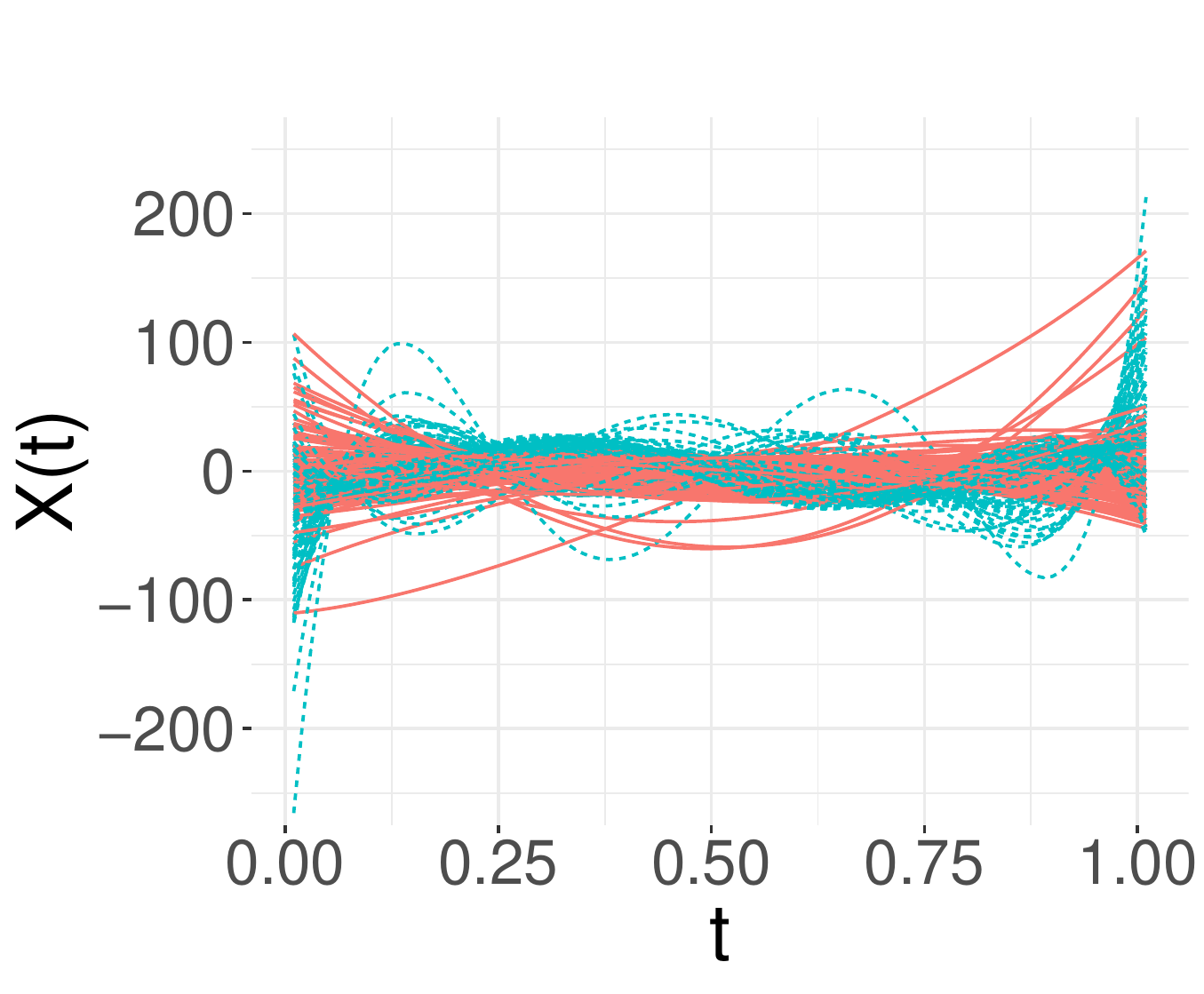}
		\caption{\centering Design \ref{case:simu2} with $\rho=10$}
	\end{subfigure}
	\caption{Samples following the simulation designs.
		Each subfigure displays a sample with
		red solid trajectories from $\Pi_0$
		and blue dashed ones from $\Pi_1$.
	}\label{fig:samples}
\end{figure}

\begin{figure}[t!]
	\centering
	\begin{subfigure}{.45\textwidth}
		\centering
		\includegraphics[width=\textwidth, height=.3\textheight]
		    {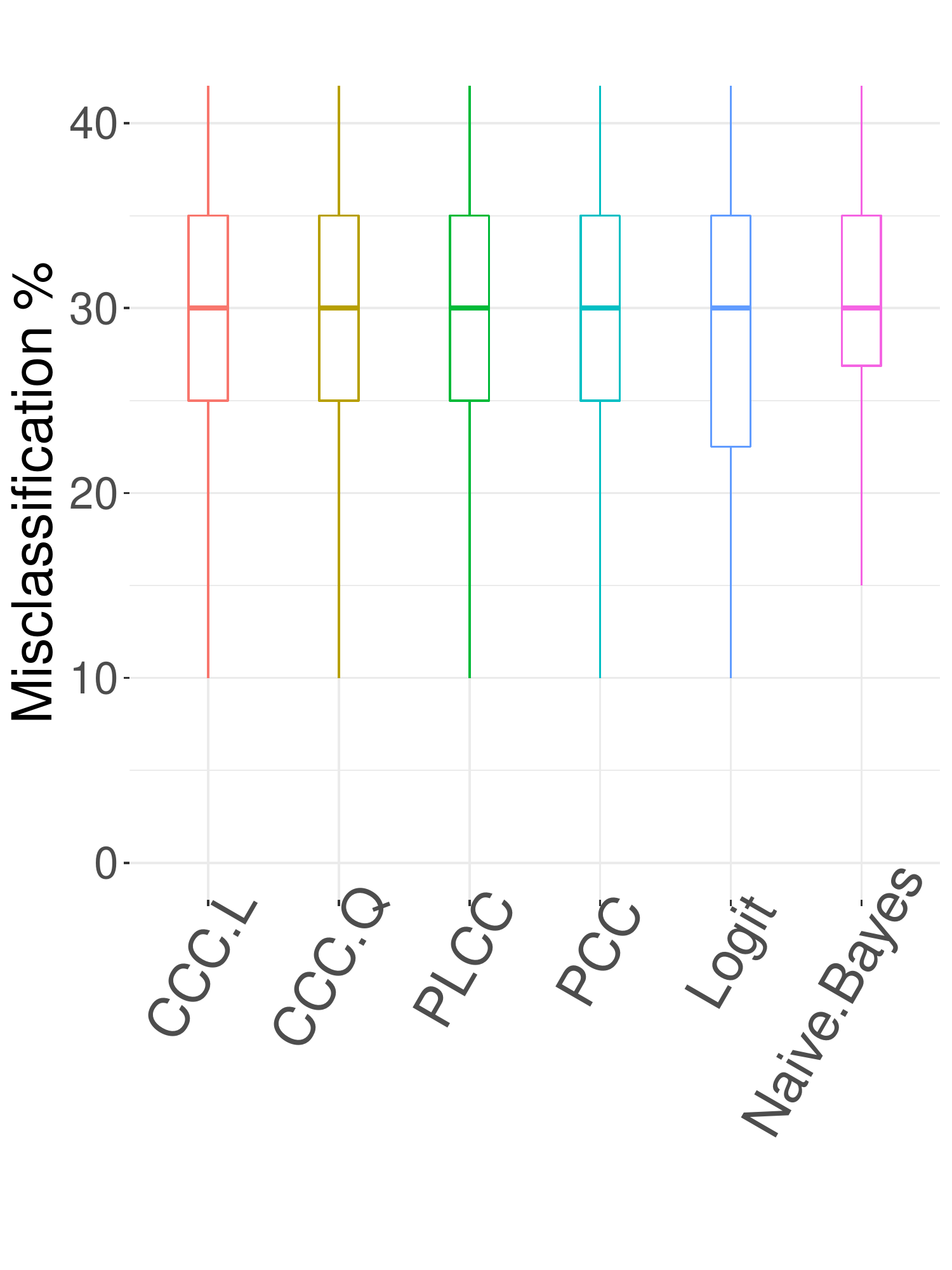}
		\caption{\centering $\rho=1$ \& $\pi_0=50\%$}
	\end{subfigure}
	\begin{subfigure}{.45\textwidth}
		\centering
		\includegraphics[width=\textwidth, height=.3\textheight]
		    {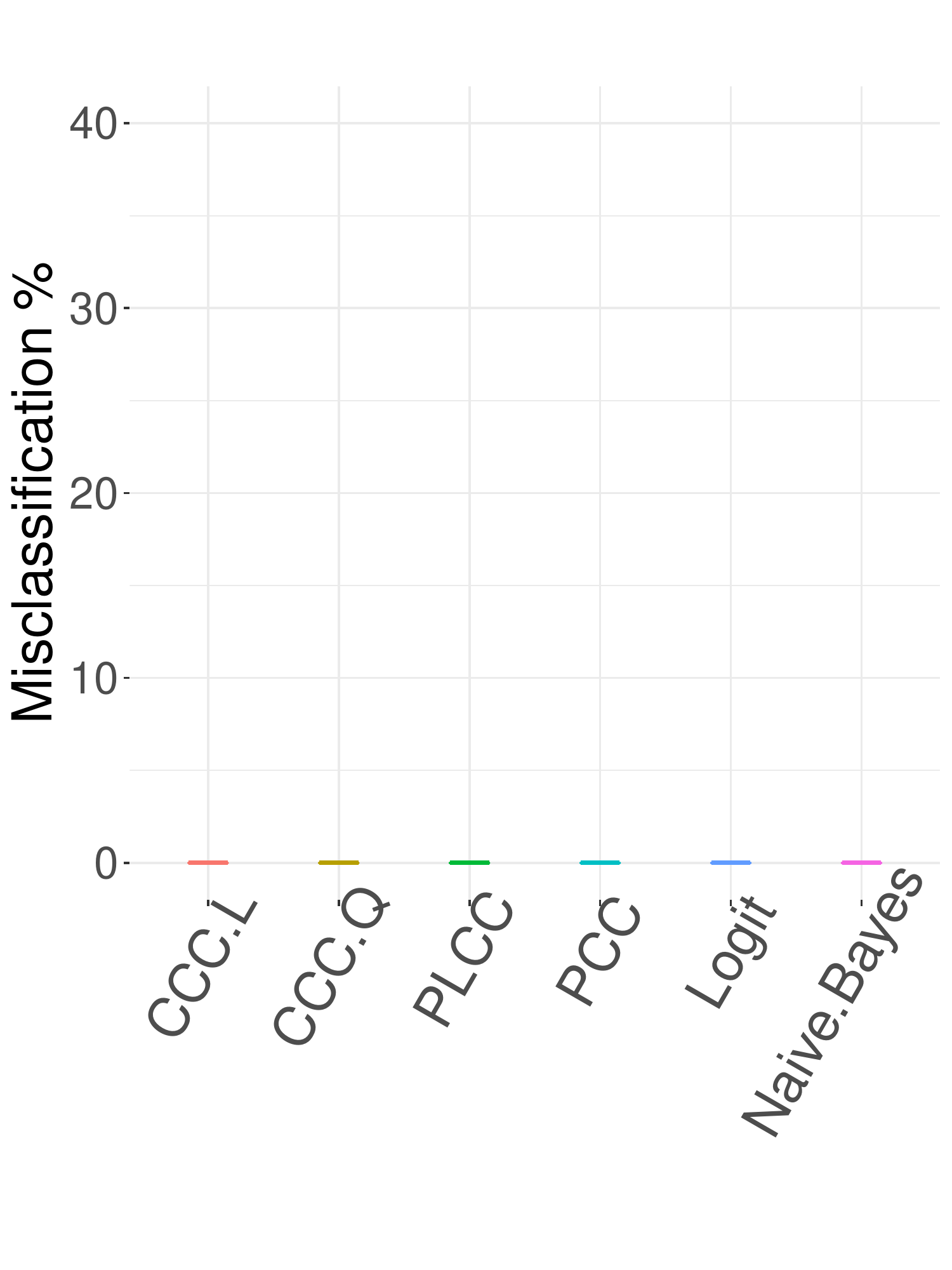}
		\caption{\centering $\rho=10$ \& $\pi_0=50\%$}
		\label{fig:simu1.10rho&50pi0}
	\end{subfigure}
	
	\begin{subfigure}{.45\textwidth}
		\centering
		\includegraphics[width=\textwidth, height=.3\textheight]
		    {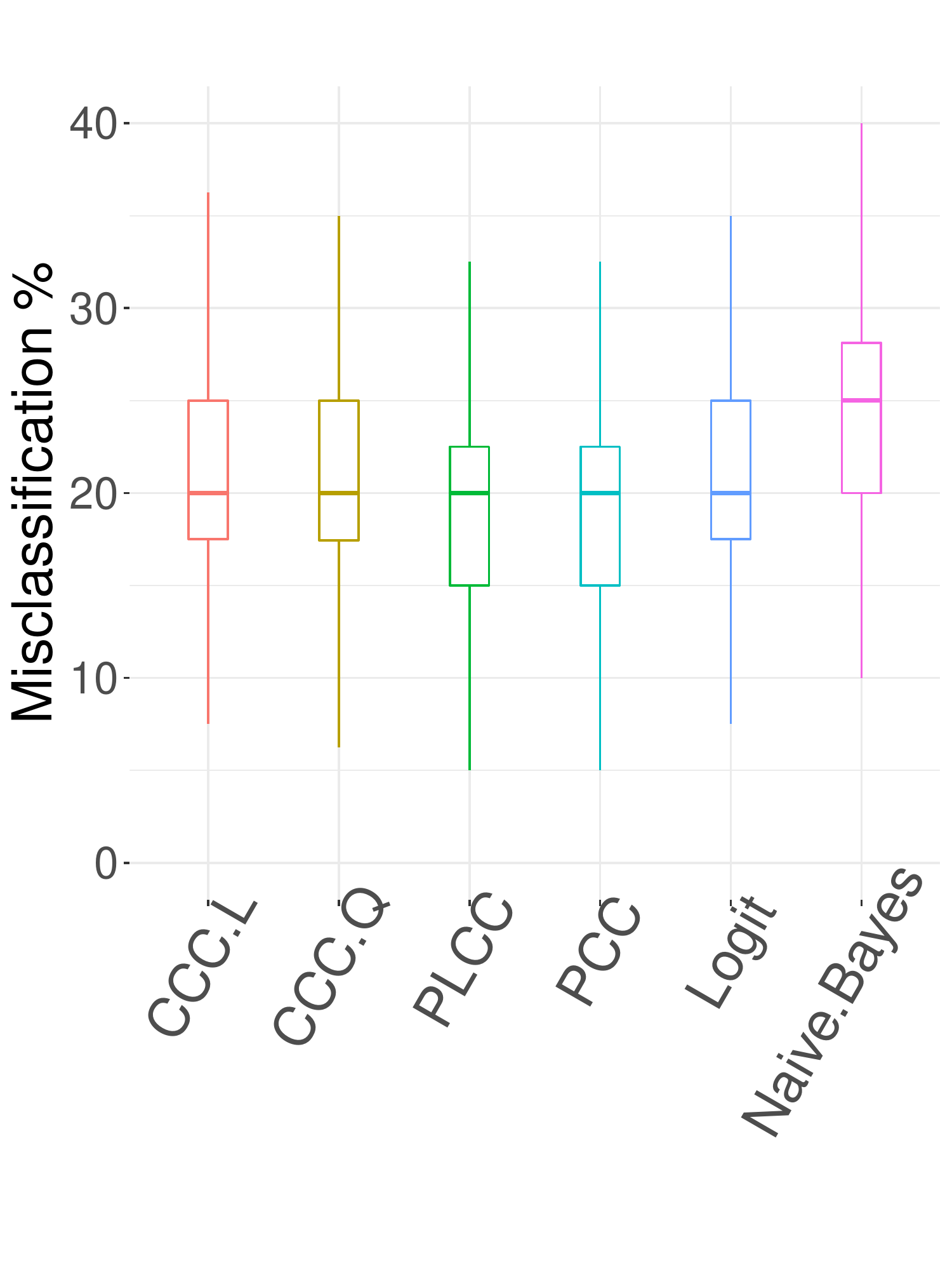}
		\caption{\centering $\rho=1$ \& $\pi_0=80\%$}
	\end{subfigure}
	\begin{subfigure}{.45\textwidth}
		\centering
		\includegraphics[width=\textwidth, height=.3\textheight]
		    {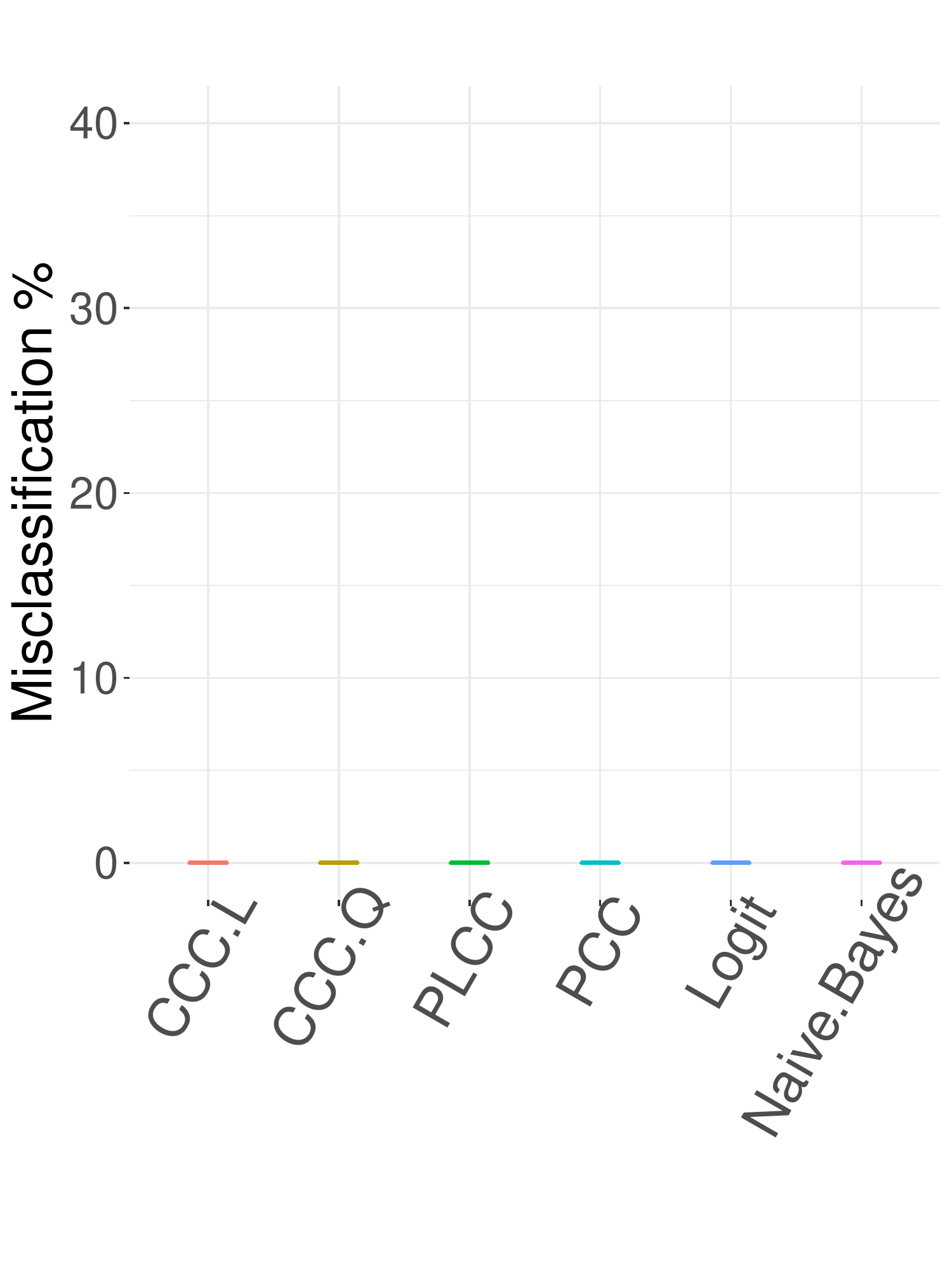}
		\caption{\centering $\rho=10$ \& $\pi_0=80\%$}
		\label{fig:simu1.10rho&80pi0}
	\end{subfigure}
	\caption{Boxplots of misclassification percentage for design \ref{case:simu1}.
		In each panel,
		the six boxes,
		from left to right,
		correspond to classifiers
		CCC-L, CCC-Q, PLCC, PCC, (functional) logit regression and (functional) naive bayes,
		respectively.
		The four subfigures come with the identical scale.
	}
	\label{fig:simu1}
\end{figure}

Regardless of values of $\rho$ and $\pi_0$,
design \ref{case:simu1} equally favored all the classifiers here (see \autoref{fig:simu1}),
as $\mu_{[1]}-\mu_{[0]}$ was parallel to the top eigenfunction of
not only $v_X$ at \eqref{eq:cov.fun.total} but also $v_X^W$ at \eqref{eq:cov.within}.
The value of $\rho$ matters a lot for design \ref{case:simu1}:
in \autoref{fig:sample.design1&rho1} (with $\rho=1$),
the two sub-populations merged together,
accompanying with extremely high misclassification rate
which may even worse than a random guess in the case of $\pi_0=80\%$
(see the second row of \autoref{tab:mean.misclassify.simu});
$\rho=10$ made the two sub-populations visibly separable
and hence corresponded to error rates of a fairly low level
(see Figures \ref{fig:simu1.10rho&50pi0} and \ref{fig:simu1.10rho&80pi0}).

\begin{figure}[t!]
	\centering
	\begin{subfigure}{.45\textwidth}
		\centering
		\includegraphics[width=\textwidth, height=.3\textheight]
		    {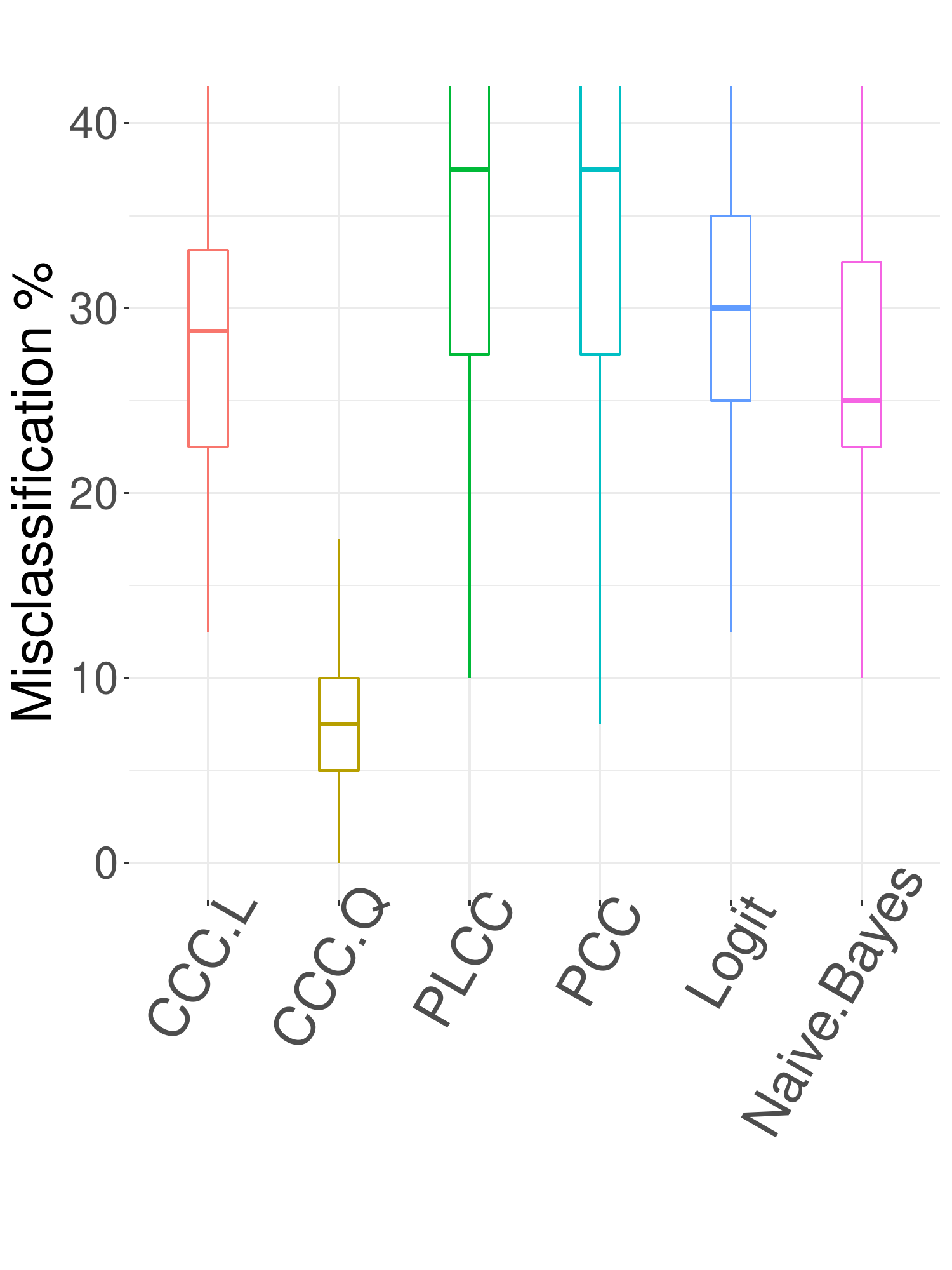}
		\caption{\centering $\rho=1$ \& $\pi_0=50\%$}
		\label{fig:simu2&rho=1&pi50}
	\end{subfigure}
	\begin{subfigure}{.45\textwidth}
		\centering
		\includegraphics[width=\textwidth, height=.3\textheight]
		    {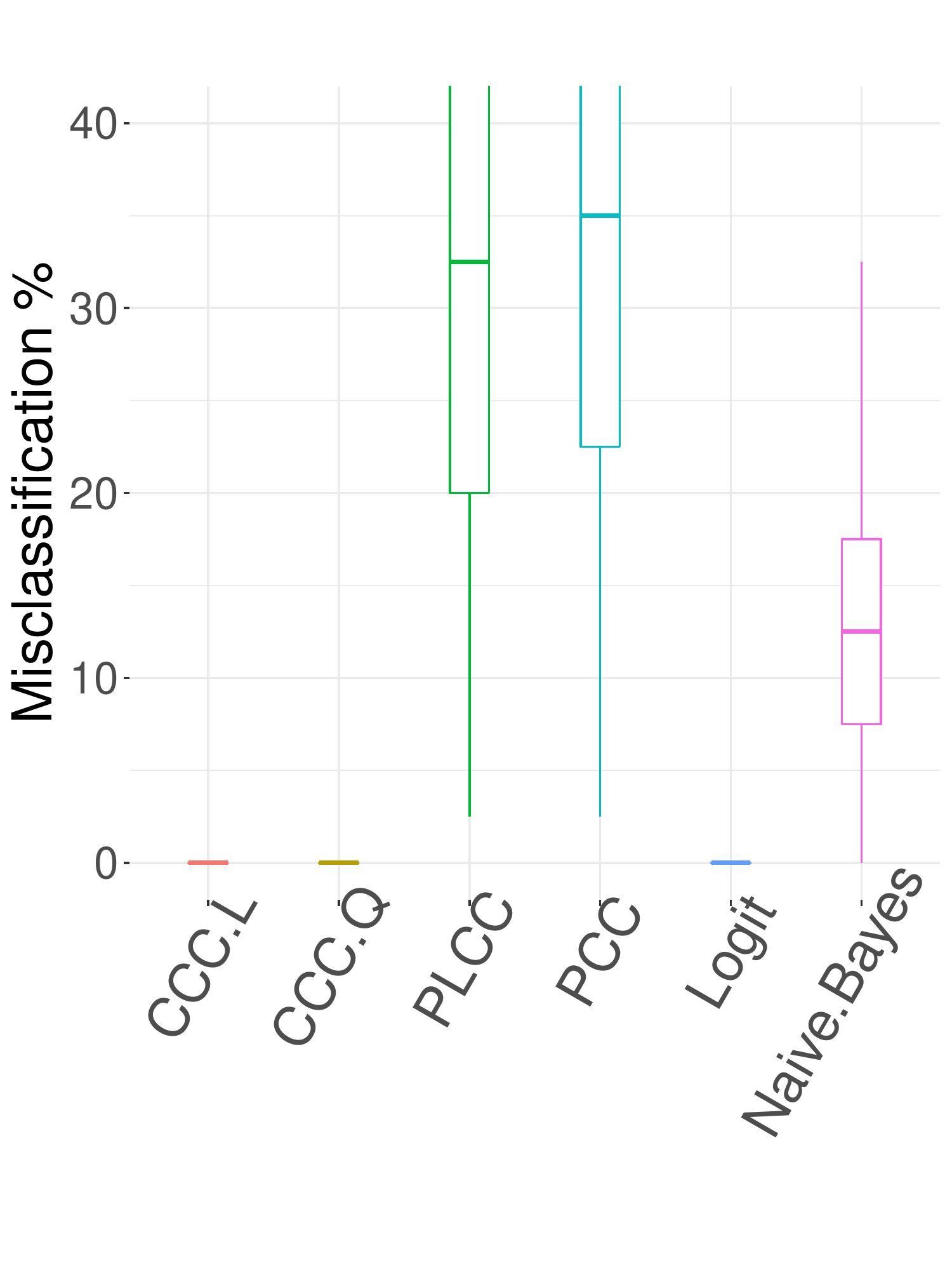}
		\caption{\centering $\rho=10$ \& $\pi_0=50\%$}
		\label{fig:simu2&rho=10&pi50}
	\end{subfigure}
	
	\begin{subfigure}{.45\textwidth}
		\centering
		\includegraphics[width=\textwidth, height=.3\textheight]
		    {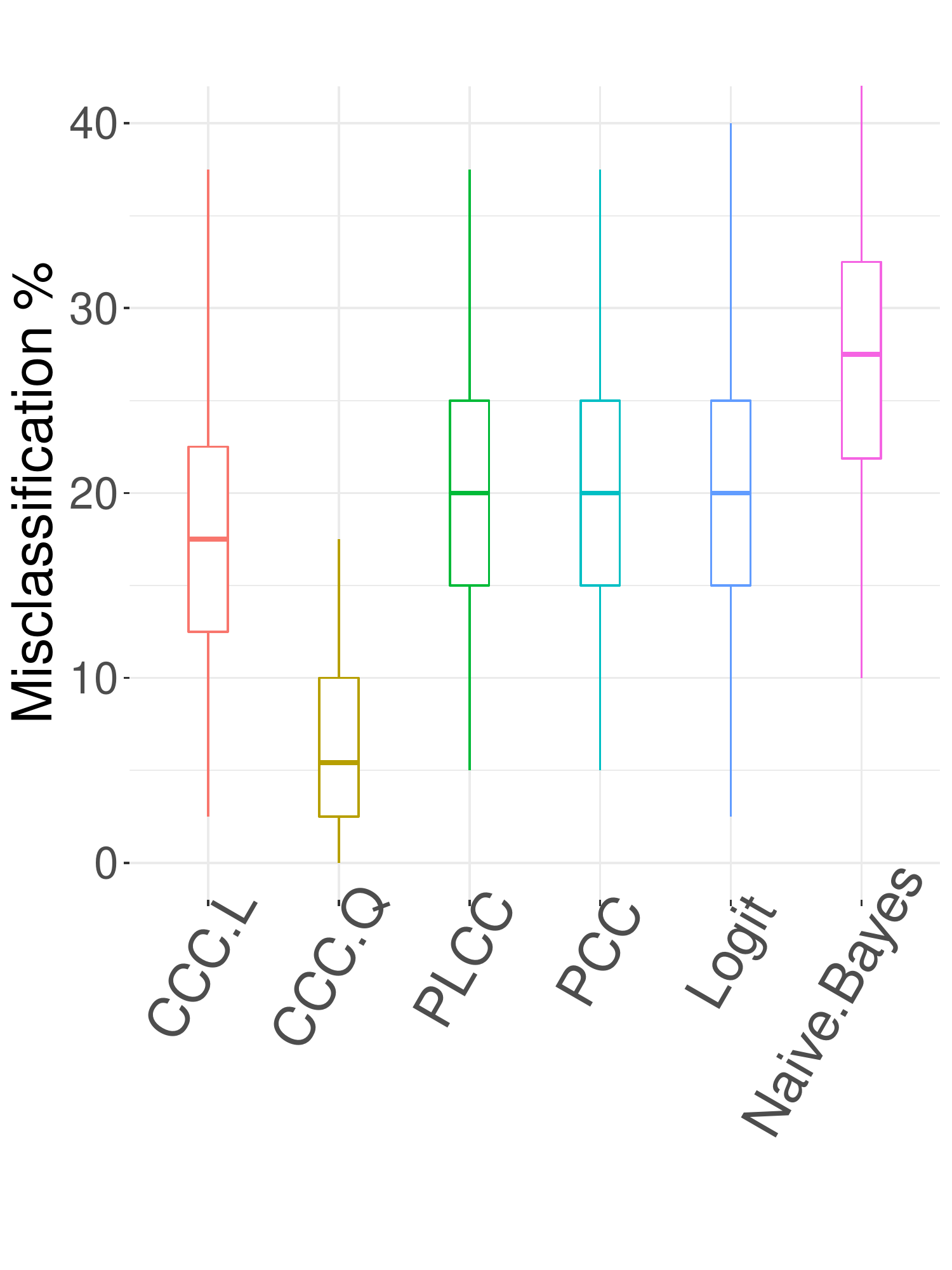}
		\caption{\centering $\rho=1$ \& $\pi_0=80\%$}
		\label{fig:simu2&rho=1&pi80}
	\end{subfigure}
	\begin{subfigure}{.45\textwidth}
		\centering
		\includegraphics[width=\textwidth, height=.3\textheight]
		    {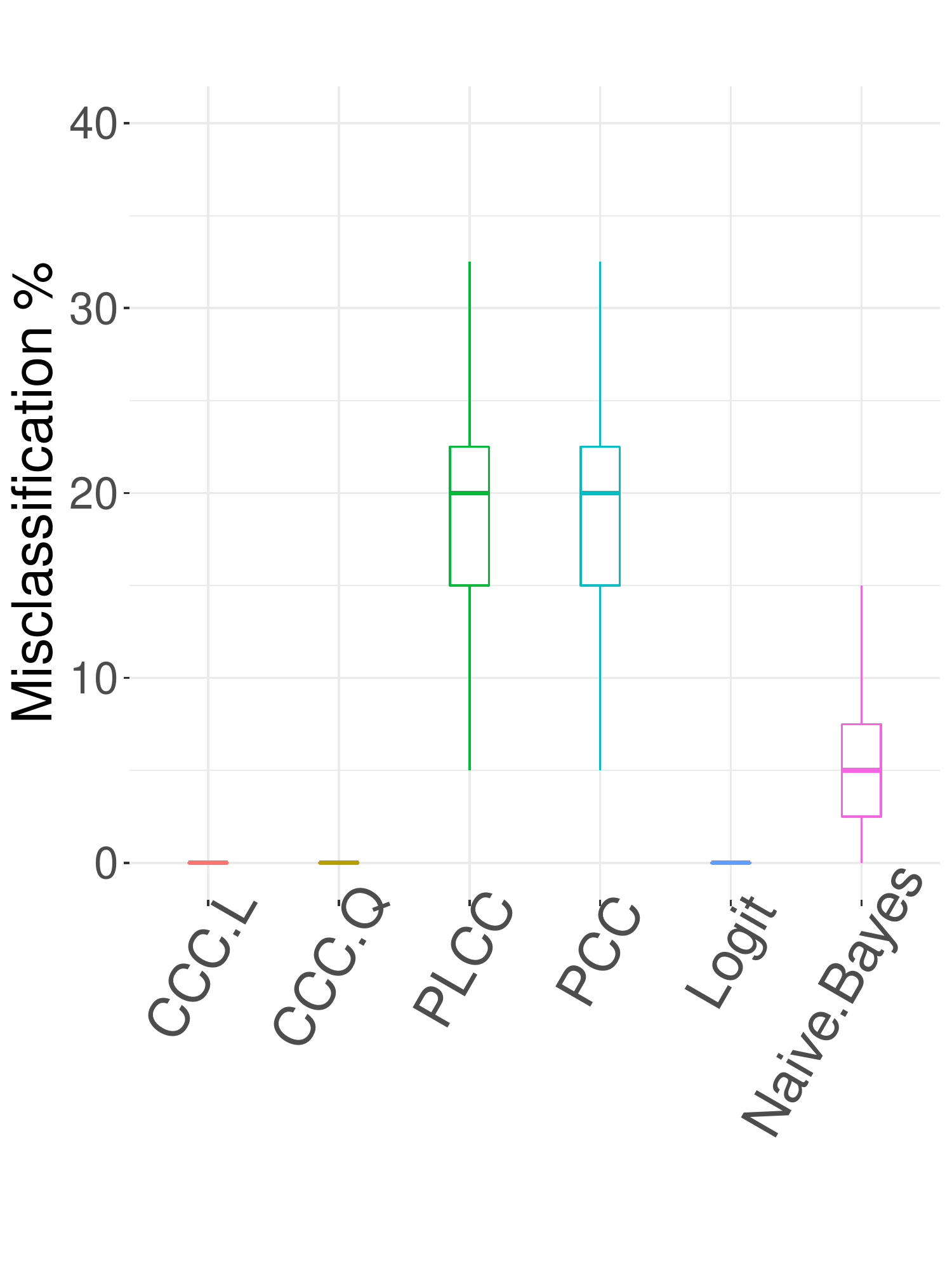}
		\caption{\centering $\rho=10$ \& $\pi_0=80\%$}
		\label{fig:simu2&rho=10&pi80}
	\end{subfigure}
	\caption{Boxplots of misclassification percentage for design \ref{case:simu2}.
		In each panel,
		the six boxes,
		from left to right,
		correspond to classifiers
		CCC-L, CCC-Q, PLCC, PCC, (functional) logit regression, and (functional) naive bayes,
		respectively.
		The four subfigures come with the identical scale.
	}
	\label{fig:simu2}
\end{figure}

It was a different story for design \ref{case:simu2}.
It restricted $\mu_{[1]}-\mu_{[0]}$ to be parallel to $\phi_{0,3}$
which is the least important eigenfunction of
\begin{multline*}
    v_X^W(s, t)
    = 160.02\phi_{0,1}(s)\phi_{0,1}(t)
    + 80.04\phi_{0,2}(s)\phi_{0,2}(t)\\
    + 40.08\phi_{0,5}(s)\phi_{0,5}(t)
    + 20.16\phi_{0,4}(s)\phi_{0,4}(t)
    + \phi_{0,3}(s)\phi_{0,3}(t).
\end{multline*}
In this case,
focused only on decomposing $v_X^W$ at \eqref{eq:cov.within},
PCC probably failed to extract the correct direction of $\mu_{[1]}-\mu_{[0]}$
and naturally yielded more misclassification
regardless of $\rho$ or $\pi_0$.
Moreover,
$v_X^{[1]}$ shared the same eigenfunctions with $v_X^{[0]}$
but in a reversed order,
violating the assumption of CCC-L and PLCC.
Due to the magnitude of $\lambda_{1,3}$,
the two subgroups appeared not separable even for $\rho=10$
(see \autoref{fig:samples}).
Actually,
trajectories from sub-population $\Pi_1$ were more bumpy;
this feature would become more obvious for larger $\rho$, 
e.g, $\rho=100$ (not illustrated here).
When $\rho=1$,
CCC-Q significantly outperformed the other competitors here;
see Figures \ref{fig:simu2&rho=1&pi50} and \ref{fig:simu2&rho=1&pi80}.
As $\rho$ grew up to 10,
the performance of classifiers was generally improved,
though PLCC and PCC still output pretty high misclassification rate
(see the last two rows of \autoref{tab:mean.misclassify.real} 
and Figures \ref{fig:simu2&rho=10&pi50} and \ref{fig:simu2&rho=10&pi80}).
If we further enlarged $\rho$ to, e.g., 100,
the identification became no longer challenging 
even for PLCC and PCC.


\begin{table}[t!]
    \centering\small
    \caption{
		Mean misclassification percentage (\%) (with standard deviations in parentheses)
		for various settings and classifiers.
		The last six columns,
		from left to right,
		correspond to classifiers
		CCC-L, CCC-Q, PLCC, PCC, (functional) logit regression and (functional) naive bayes,
		respectively.
	}\label{tab:mean.misclassify.simu}
    \begin{tabular}{ccccccccc}
        \hline
        Design
        & $\rho$
        & $\pi_0$
        & CCC-L
        & CCC-Q
        & PLCC
        & PCC
        & Logit
        & Naive Bayes
        \\\hline
        \ref{case:simu1}
        & 1
        & 50\%
        & 30 (8.2)
        & 30 (8.6)
        & 31 (7.1)
        & 31 (7.3)
        & 30 (8.0)
        & 31 (7.9)
        \\
        \ref{case:simu1}
        & 1
        & 80\%
        & 21 (5.9)
        & 21 (6.0)
        & 20 (5.9)
        & 20 (6.0)
        & 21 (5.9)
        & 24 (6.6)
        \\
        \ref{case:simu1}
        & 10
        & 50\%
        & .13 (.56)
        & .15 (.60)
        & .19 (1.0)
        & .11 (.52)
        & .09 (.46)
        & .15 (.60)
        \\
        \ref{case:simu1}
        & 10
        & 80\%
        & .22 (.70)
        & .22 (.72)
        & .31 (1.1)
        & .25 (.75)
        & .11 (.52)
        & .24 (.78)
        \\\hline
        \ref{case:simu2}
        & 1
        & 50\%
        & 29 (7.8)
        & 7.4 (4.1)
        & 38 (14)
        & 37 (14)
        & 29 (8.0)
        & 27 (7.9)
        \\
        \ref{case:simu2}
        & 1
        & 80\%
        & 17 (7.0)
        & 6.7 (4.0)
        & 20 (6.1)
        & 20 (6.2)
        & 20 (6.8)
        & 27 (8.3)
        \\
        \ref{case:simu2}
        & 10
        & 50\%
        & .22 (.73)
        & .21 (.62)
        & 33 (15)
        & 35 (15)
        & .11 (.52)
        & 13 (6.9)
        \\
        \ref{case:simu2}
        & 10
        & 80\%
        & .23 (.70)
        & .25 (.79)
        & 20 (6.3)
        & 20 (6.7)
        & .14 (.57)
        & 5.3 (4.4)
        \\\hline
    \end{tabular}
\end{table}

\subsection{Real data application}

For each dataset,
we repeated a random split with ratio 8:2 for 200 times:
each time we trained classifiers with 80\% data points
and then tested them on the remaining 20\%.
Table \ref{tab:mean.misclassify.real} summarized
the means and standard deviations of misclassification percentages.

\begin{figure}[t!]
	\centering
	\begin{subfigure}{.45\textwidth}
		\centering
		\includegraphics[width=\textwidth, height=.3\textheight]
		    {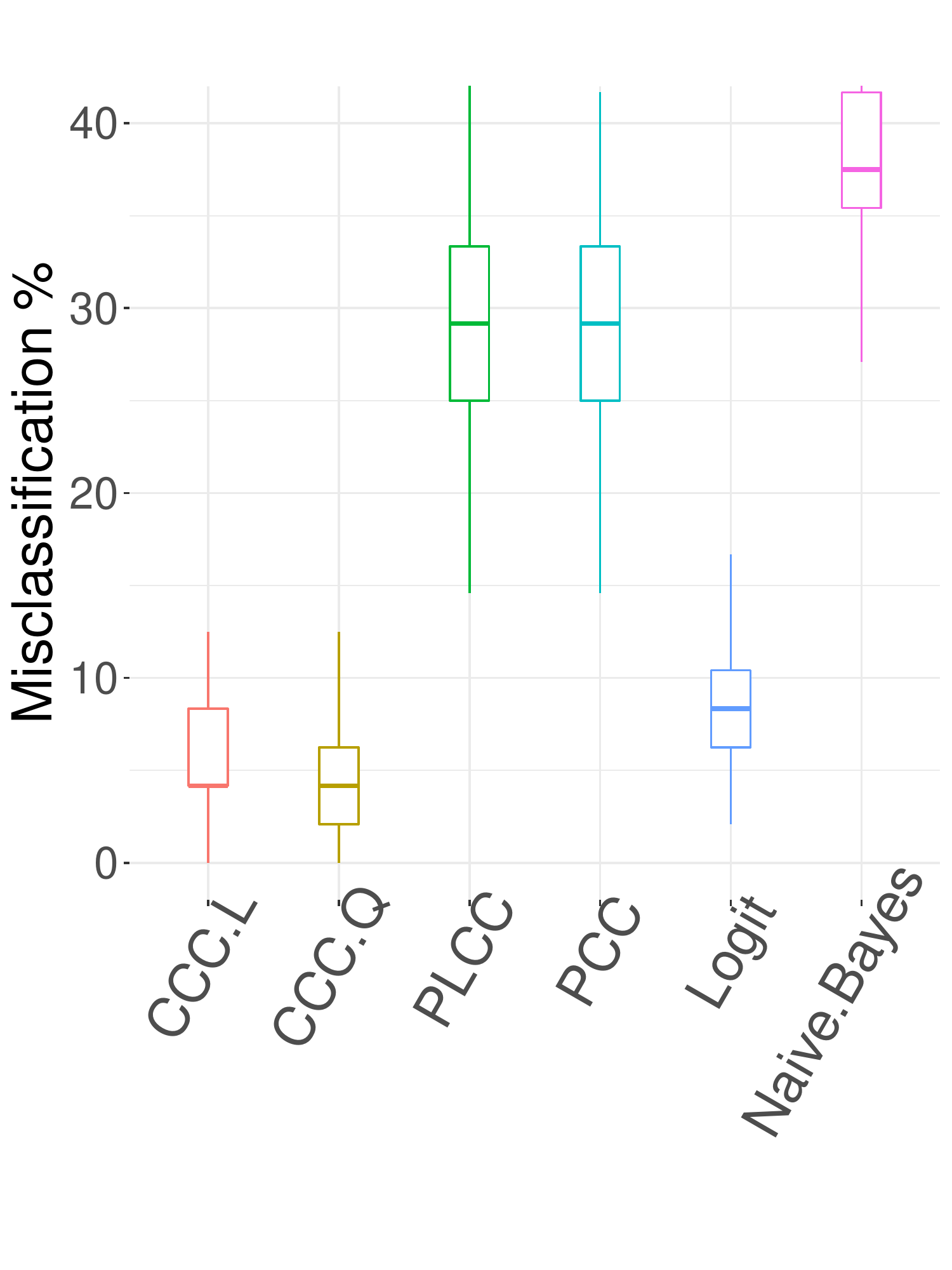}
		\caption{\centering Tecator\texttrademark\ (original)}
		\label{fig:tecator.original}
	\end{subfigure}
	\begin{subfigure}{.45\textwidth}
		\centering
		\includegraphics[width=\textwidth, height=.3\textheight]
		    {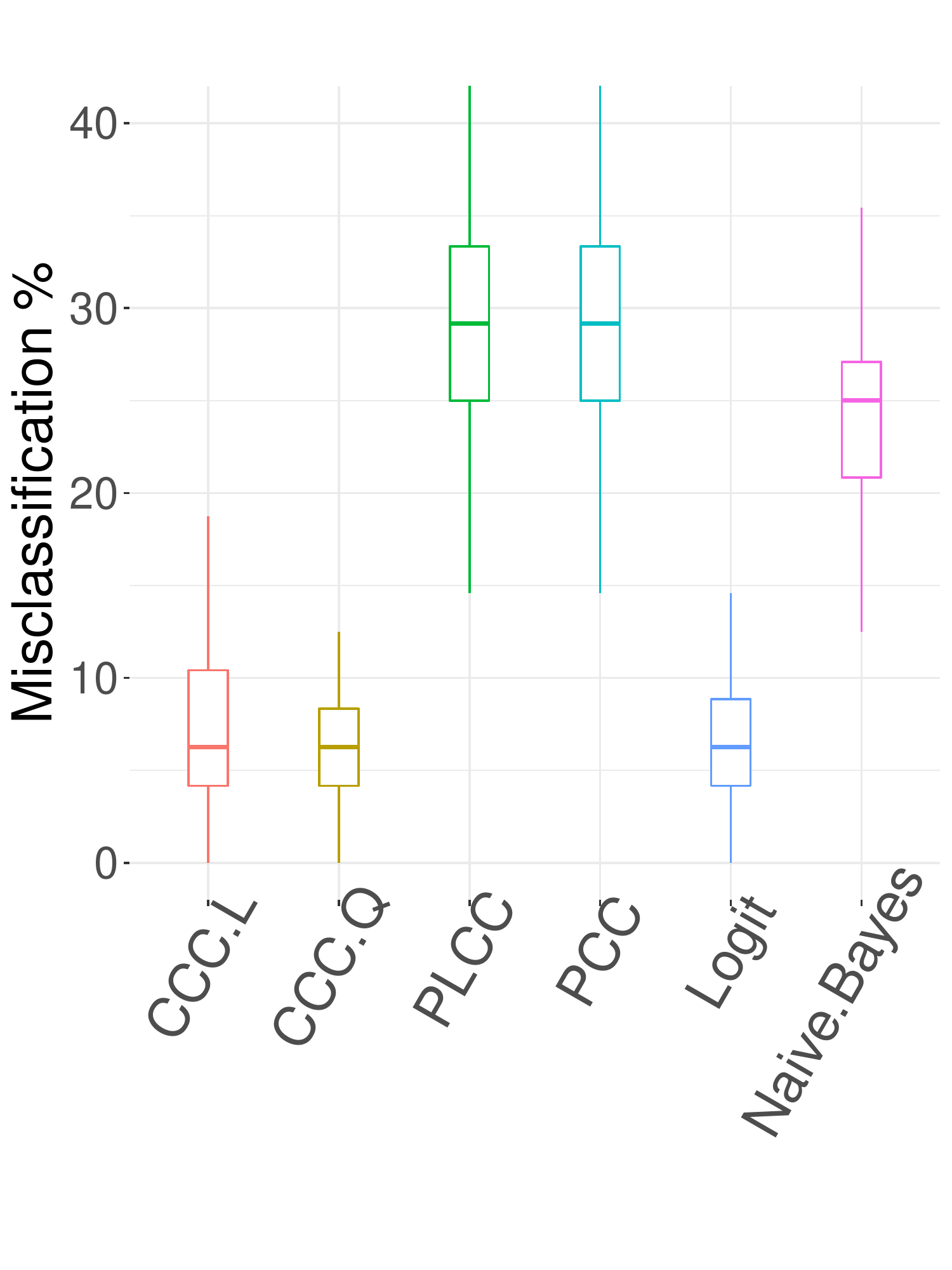}
		\caption{\centering Tecator\texttrademark\ (2nd order derivative)}
		\label{fig:tecator.derivative}
	\end{subfigure}
	
	\begin{subfigure}{.45\textwidth}
		\centering
		\includegraphics[width=\textwidth, height=.3\textheight]
		    {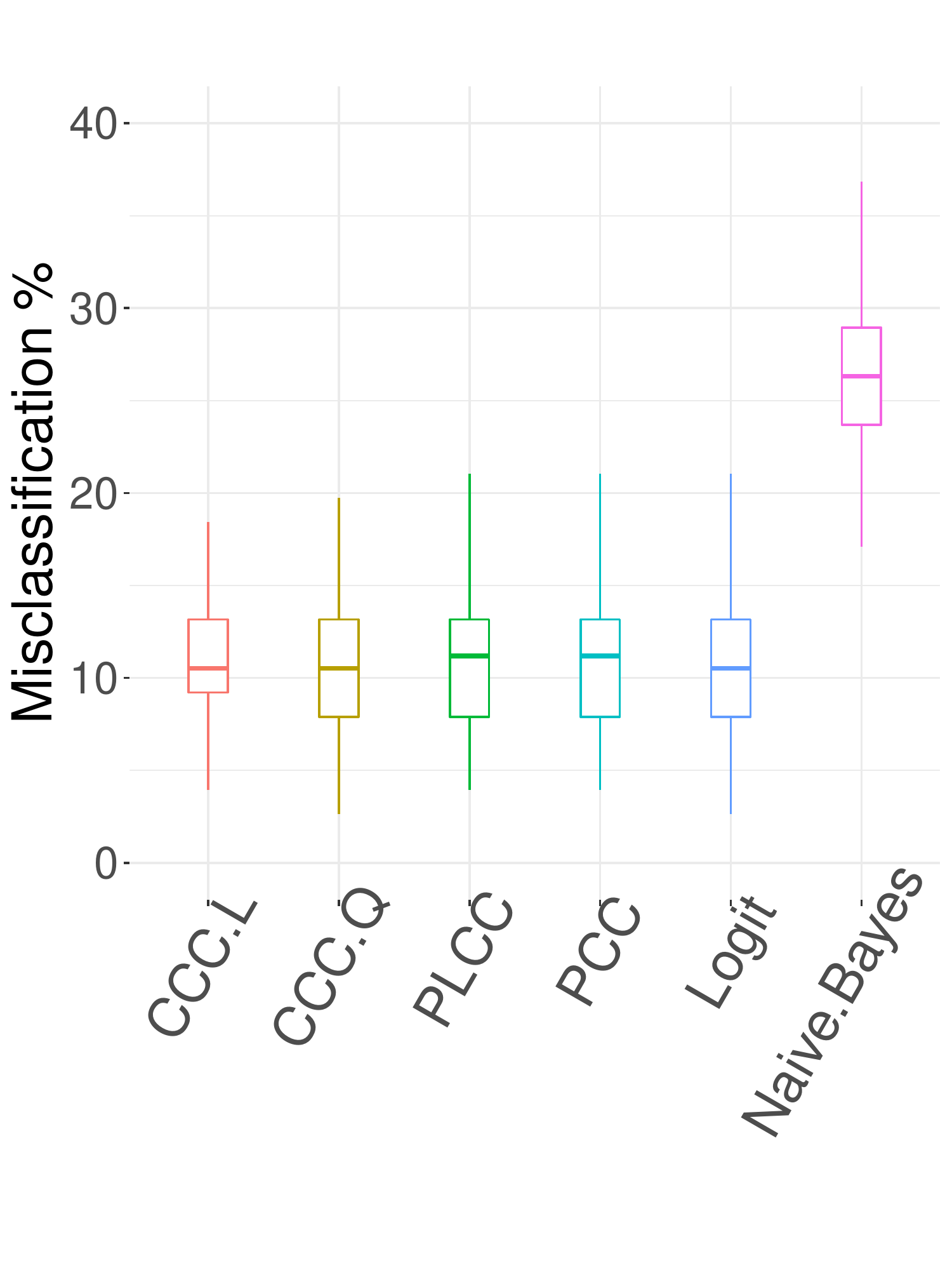}
		\caption{\centering DTI}\label{fig:dti}
	\end{subfigure}
	\caption{Boxplots of misclassification percentages for
	    Tecator\texttrademark\ and DTI data.
	    The top two panels reflect respective results with
	    original and second order derivative curves.
		In each panel,
		the four boxes,
		from left to right,
		correspond to classifiers
		CCC-L, CCC-Q, PLCC, PCC, (functional) logit regression, and (functional) naive bayes,
		respectively.
		The three subfigures are displayed with the identical scale.
	}
	\label{fig:real}
\end{figure}

The first real example treated Tecator\texttrademark~data
(accessible at \url{http://lib.stat.cmu.edu/datasets/tecator} on \today).
This dataset was collected by
Tecator\texttrademark~Infratec Food and Feed Analyzer
(and hence named).
It consisted of near infrared absorbance spectra
(i.e., the logarithm to base 10 of transmittance at each wavelength)
of 240 fine-chopped pure meat samples.
Each spectrum ranged from 850 to 1050 nm
and was spotted at 100 ``time points'' (viz. channels).
Additionally, three contents, water, protein and fat,
were recorded in percentage for each piece of meat.
In our study,
meat samples were categorized into two groups:
$\Pi_1$ was comprised of meat samples with protein content less than 16\%
and the rest constituted $\Pi_0$.
The 240 spectrum curves
(or their second order derivative curves
as recommended by \citealp[][Section 7.2.2]{FerratyVieu2006})
were regarded as functional covariates in the study.
For both sorts of trajectories,
the classification output was analogous to 
that for simulated design \ref{case:simu2} with $\rho=10$:
CCC subtypes and (functional) logit regression yielded 
considerably less errors than the other three classifiers;
compare Figures \ref{fig:simu2&rho=10&pi50}, \ref{fig:simu2&rho=10&pi80}, 
\ref{fig:tecator.original} and \ref{fig:tecator.derivative}.

We returned to DTI mentioned in \autoref{sec:introduction}.
Measured through DTI,
the fractional anisotropy,
a scalar ranging from 0 to 1,
is used to reflect the fiber density, axonal diameter and myelination in white matter.
Along a tract of interest,
fractional anisotropy values form a curve, viz. a tract fractional anisotropy profile.
We analyzed dataset \texttt{DTI} in R-package \texttt{refund} \cite[][]{R-refund}.
It was initially collected at Johns Hopkins University and the Kennedy-Krieger Institute,
containing tract fractional anisotropy profiles 
(measured at 93 locations)
for corpus callosum of
healthy people ($\Pi_0$) and MS patients ($\Pi_1$) 
and involving 382 subjects in total.
By classifying these profiles 
(with missing values imputed through local polynomial regression),
we tried to identify the status of each subject:
healthy or suffering from MS.
Except for the (functional) naive bayes,
classifiers reached a tie after rounding 
(see the last row of \autoref{tab:mean.misclassify.real}),
i.e., 
they enjoyed fairly close classification accuracy in identifying MS patients.

\begin{table}[t!]
    \centering
    \caption{
		Average misclassification percentage for real datasets
		(with standard deviations in parentheses)
		corresponding to different classifiers.
		The last six columns correspond to classifiers
		CCC-L, CCC-Q, PLCC, PCC, (functional) logit regression, and (functional) naive bayes,
		respectively.
		Ties at the last row are caused by rounding.
	}\label{tab:mean.misclassify.real}
	\begin{tabular}{@{}ccccccc@{}}
		\hline
		\multicolumn{1}{l}{}           
		& CCC-L     
		& CCC-Q    
		& PLCC     
		& PCC      
		& Logit     
		& Naive Bayes 
		\\\hline
		Tecator\texttrademark\ (original)             
		& 5.5 (3.3) 
		& 4.6 (2.8) 
		& 30 (6.5) 
		& 29 (6.8) 
		& 8.8 (4.0) 
		& 39 (6.6)    
		\\
		Tecator\texttrademark\ (2nd order derivative) 
		& 7.3 (4.4) 
		& 6.0 (3.4) 
		& 30 (6.5) 
		& 30 (6.5) 
		& 7.0 (3.6) 
		& 24 (5.9)    
		\\
		DTI                            
		& 11 (3.5)  
		& 11 (3.5)  
		& 11 (3.6) 
		& 11 (3.6) 
		& 11 (3.7)  
		& 26 (4.5)    
		\\\hline
	\end{tabular}
\end{table}

\section{Conclusion and discussion}\label{sec:conclusion}

We propose two subtypes of CCC classifiers,
viz. CCC-L and -Q,
for binary classification of curves.
Theoretically,
under certain circumstances,
CCC-L enjoys the (asymptotic) zero misclassification
regardless of the distribution assumption,
while,
in certain empirical studies,
CCC-Q seems superior to CCC-L and other competitors.
Once regularity conditions are met,
our proposal results in empirical classifiers
which are consistent to their theoretical counterparts,
for the case of ``fixed $p$ and infinite $N$''.

In the numerical experiments in \autoref{sec:numerical},
we do see benefits from the introduction of the supervision controller $\alpha$.
Nevertheless,
one cannot be too optimistic;
actually,
tuning one more parameter may yield more variation and even bias.
In addition,
the projection direction $\beta_{p,\alpha}$ at \eqref{eq:beta.p.alpha} potentially becomes unstable for large $p$,
since $w_{p,\alpha}$ is constructed iteratively without penalty.
Nevertheless,
the introduction of one more hyperparameter
would force our implementation to be more computationally involved.

One may be concerned about the chosen values of hyperparameters for our proposed classifiers.
In most cases of numerical study,
hyperparameter $\alpha$ (resp. $p$) for CCC-L and -Q ended with similar values;
exceptions arose for design \ref{case:simu2} with $\rho=1$
(where CCC-Q outperformed CCC-L significantly).
Specifically,
in that case,
CCC-Q picked up mean $\alpha$ around .7,
whereas CCC-L chose about .5 and hence yielded a classification accuracy comparable with PLCC. 
As for the size of basis,
CCC subtypes seemed even more ``parsimonious'' than PLCC,
mostly utilizing less than 3 basis functions.

Regarding a further extension to the functional classification with $K$ ($\geq 3$) classes,
one naive strategy is to carry out binary classifiers repeatedly.
To be explicit,
for a newcomer $X^*$,
each time we only consider two distinct labels
and then assign either label to it,
or equivalently, throw a vote for either of the two labels.
After all $\genfrac(){0pt}{2}{K}{2}$ binary classifications,
the label that wins the most votes is eventually assigned to $X^*$.
Another route is embedding FC basis into
the iteratively reweighted least squares \citep{Green1984}.
In that way,
the resulting modification could be utilized 
for estimating generalized linear models with functional covariates.

\newcommand{\newblock}{}
\bibliographystyle{cjs}
\bibliography{mybibfile}

\begin{appendix}

Our theoretical perspectives are built upon the following assumptions. 
\begin{enumerate}[label=(C\arabic*)]
	\item\label{cond:unbounded}
	    $
	        \{\int_{\mathcal{T}}\beta_{p, \alpha}(\mu_{[1]}-\mu_{[0]})\}^2/
	        \var\{\int_{\mathcal{T}}\beta_{p, \alpha}(X-\mu_{[k]})\mid X\in\Pi_k\}
	    $
	    diverges
        as $p\to\infty$ for each $\alpha$ and $k$.
    \item\label{cond:x.derivative}
        Realizations of $X$ is twice continuously differentiable
        and $\|X'\|$ is bounded almost surely.
    \item\label{cond:eigenvalue}
        $\tau_1,\ldots,\tau_L$ are eigenvalue of
        $\mathbf{Pen}^{-1/2}\mathbf{W}\mathbf{Pen}^{-1/2}$
        such that
        $\tau_1=\tau_2=0$,
        $\tau_3\geq\cdots\geq\tau_L$,
        and
        $C_1(l-2)^{-4}\leq \tau_l\leq C_2(l-2)^{-4}$ for $l\geq 3$,
        with neither $C_1$ nor $C_2$ depending on $l$ or $L$.
    \item\label{cond:theta}
        $M\to\infty$ and
        $M^{-1}\theta_0\to 0$
        as $N\to\infty$,
        where $\theta_0>0$ is the smoothing parameter for all trajectories.
    \item\label{cond:unique}
        For all $j$ ($\leq p\leq{\rm rank}(\hat{\mathbf{C}}_{\rm c}\mathbf{W}^{1/2})$),
        $T_{j,\alpha}(w)$
        attains a unique maximizer (up to sign) in $\{w\in L^2(\mathcal{T}):\|w\|=1\}$.
\end{enumerate}

Condition \ref{cond:unbounded} implies that,
after projected to the direction of $\beta_{p,\alpha}$,
as $p$ diverges,
the within-group covariance becomes more and more ignorable
when compared with the between-group one,
i.e., the two groups become more and more separable.
It is analogous to assumption (4.4)(d) of \citet{DelaigleHall2012}
and assures us of the (asymptotic) perfect classification of CCC-L.
Assumptions \ref{cond:x.derivative} and \ref{cond:eigenvalue}
jointly guarantee that
the smoothed curves converge to the true ones as observations become denser and denser;
although the latter one has been proved by
\citet[][ eq. 4]{Utreras1983} for natural splines,
we have little knowledge on whether it still holds for B-splines
and hence have to assume it following \citet[][ eq. A4.3.1]{CravenWahba1979}.
If we have extra regularity conditions \ref{cond:theta} and \ref{cond:unique},
the proposed empirical implementation in \autoref{sec:implementation}
turns out to be consistent in probability.

\begin{proof}{Proof of \autoref{prop:0.error}}{}
    Write
    $
        \gamma_{p, \alpha}
        =\int_{\mathcal{T}}
            \beta_{p, \alpha}(\mu_{[1]}-\mu_{[0]})
    $
    and
    $
        R_{p, \alpha}^{[k]}
        =\int_{\mathcal{T}}
            \beta_{p, \alpha}(X^*-\mu_{[k]})
    $.
    Recalling \eqref{eq:sigma.k},
    $
        \sigma_{[k]}^2(\beta_{p, \alpha})
        =\var(R_{p, \alpha}^{[k]}\mid X\in\Pi_k)
    $,
    $k=0,1$.
    Thus,
    \begin{align*}
        \Pr\{&\mathcal{D}_L(X^*\mid\beta_{p, \alpha})<0\mid X^*\in\Pi_0\}
        \\
        &= \Pr\left\{
                (R_{p, \alpha}^{[0]}-\gamma_{p, \alpha})^2-(R_{p, \alpha}^{[0]})^2
                <2\sigma_{[0]}^2(\beta_{p, \alpha})\ln\frac{1-\pi_0}{\pi_0}
                \,\middle|\, X^*\in\Pi_0
            \right\}
        \\
        &=\Pr\left[\frac{R_{p, \alpha}^{[0]}}{\sigma_{[0]}(\beta_{p, \alpha})}>
                            \frac{\gamma_{p, \alpha}^2+
                                2\sigma_{[0]}^2(\beta_{p, \alpha})
                                    \ln\{\pi_0/(1-\pi_0)\}
                                }
                                {2\gamma\sigma_{[0]}(\beta_{p, \alpha})}
                        \,\middle|\, X^*\in\Pi_0\right]
        \\
        &\leq \frac{4\sigma_{[0]}^2(\beta_{p, \alpha})/\gamma_{p, \alpha}^2}
            {\left[1+
                2\gamma_{p, \alpha}^{-2}
                    \sigma_{[0]}^2(\beta_{p, \alpha})\ln\{\pi_0/(1-\pi_0)\}
            \right]^2},
    \end{align*}
    where the upper bound is derived from Chebyshev's inequality
    and the identity that
    $R_{p, \alpha}^{[0]}/\sigma_{[0]}(\beta_{p, \alpha})$
    (conditional on the event $X^*\in\Pi_0$)
    is of zero mean and unit variance.
    Similarly,
    we deduce that
    $$
        \Pr\{\mathcal{D}_L(X^*\mid\beta_{p, \alpha})>0\mid X^*\in\Pi_1\}
        \leq \frac{4\sigma_{[1]}^2(\beta_{p, \alpha})/\gamma_{p, \alpha}^2}
            {\left[1+
                2\gamma_{p, \alpha}^{-2}\sigma_{[0]}^2(\beta_{p, \alpha})
                    \ln\{(1-\pi_0)/\pi_0\}
            \right]^2}.
    $$
    Eventually,
    as $p$ diverges,
    the zero-convergence of
    \begin{multline*}
        {\rm err}\{\mathcal{D}_L(X^*\mid\beta_{p, \alpha})\}
        =\pi_0\Pr\{\mathcal{D}_L(X^*\mid\beta_{p, \alpha})<0\mid X^*\in\Pi_0\}
        \\
        +(1-\pi_0)\Pr\{\mathcal{D}_L(X^*\mid\beta_{p, \alpha})>0\mid X^*\in\Pi_1\}
    \end{multline*}
    results from \ref{cond:unbounded}
    (i.e., $\sigma_{[k]}^2(\beta_{p, \alpha})/\gamma_{p, \alpha}^2\to 0$
    as $p$ diverges for each $\alpha$ and $k$).
\end{proof}

\begin{proof}{Proof of \autoref{prop:x.i}}{}
    Recall $\Delta t = (t_{\max}-t_{\min})/M$
    and matrices
    $\ppsi$ \eqref{eq:ppsi},
    $\hat{\bm{c}}_i$ \eqref{eq:c.i.hat},
    $\PPsi$ \eqref{eq:PPsi},
    $\mathbf{W}$ \eqref{eq:W}
    and $\mathbf{Pen}$ \eqref{eq:Pen},
    all defined in Section \ref{sec:implementation}.
    Introduce operator $\mathcal{P}_{BS_L}$
    such that,
    for each $f\in L^2(\mathcal{T})$,
    $\mathcal{P}_{BS_L}f$ is the orthogonal projection of $f$ onto $BS_L$ of \eqref{eq:bs.k},
    i.e.,
    $$
        \mathcal{P}_{BS_L}f
        =\left[\int_{\mathcal{T}}f\psi_1,\ldots,\int_{\mathcal{T}}f\psi_L\right]
            \mathbf{W}^{-1}\ppsi.
    $$
    For each $i$,
    specifically,
    $\mathcal{P}_{BS_L}X_i = c_i^\top\ppsi$
    with
    $$
        c_i
        =\mathbf{W}^{-1}\left[\int_{\mathcal{T}}X_i\psi_1,\ldots,\int_{\mathcal{T}}X_i\psi_L\right]^\top.
    $$
    We chop $\|\hat{X}_i-X_i\|$ into two segments:
    $\|\mathcal{P}_{BS_L}X_i-X_i\|$ and $\|\hat{X}_i-\mathcal{P}_{BS_L}X_i\|$.
    Combined with \citet[][ Theorem 16]{LycheManniSpeleers2018},
    condition \ref{cond:x.derivative} implies
    $$
        \|\mathcal{P}_{BS_L}X_i-X_i\|
        =O_p(\Delta t)
        =O_p(M^{-1})
        \quad\text{as}\quad
        M\to\infty.
    $$

    Further,
    condition \ref{cond:x.derivative} allows us to
    follow \citet[][ Theorem 5]{Chui1971} to verify that,
    as $M\to\infty$,
    $\|\mathbf{W}-\Delta t\PPsi^\top\PPsi\|_F^2 = O(M^{-2})$,
    $\|\mathbf{W}-\Delta t\PPsi^\top\PPsi-\Delta t\theta_0\mathbf{Pen}\|_F^2 = O(1)$,
    and
    $$
    	\left\|
    	    \left[\int_{\mathcal{T}}X_i\psi_1,\ldots,\int_{\mathcal{T}}X_i\psi_L\right]^\top
    	    -\Delta t\bm{X}_i^\top\PPsi
    	\right\|_F^2
    	= O_p(M^{-3}),
    $$
    where $\|\cdot\|_F$ denotes the Frobenius norm.
    Let $\tau_1,\ldots,\tau_L$ be eigenvalues of
    $$
        \mathbf{Z}=\mathbf{Pen}^{-1/2}\mathbf{W}\mathbf{Pen}^{-1/2}
    $$
    with corresponding eigenvectors $\bm{e}_1,\ldots,\bm{e}_L$.
    Note the finiteness of both
    $\lim_{L\to\infty}\max_{l_1,l_2}|\int_{\mathcal{T}}\psi_{l_1}\psi_{l_2}|$
    and $\lim_{L\to\infty}\max_{l_1,l_2}|\int_{\mathcal{T}}\psi_{l_1}''\psi_{l_2}''|$.
    Thus the squared second trunk is
    \begin{align*}
        \|&\hat{X}_i-\mathcal{P}_{BS_L}X_i\|^2
        \\
        &=(\hat{\bm{c}}_i^\top-\bm{c}_i^\top)\mathbf{W}(\hat{\bm{c}}_i-\bm{c}_i)
        \\
        &=\left\{
            \Delta t\bm{X}_i^\top\PPsi(\Delta t\PPsi^\top\PPsi+\Delta t\theta_0\mathbf{Pen})^{-1}
            -\bm{c}_i^\top
        \right\}
        \mathbf{W}
        \left\{
            (\Delta t\PPsi^\top\PPsi+\Delta t\theta_0\mathbf{Pen})^{-1}\Delta t\PPsi^\top\bm{X}_i
            -\bm{c}_i
        \right\}
        \\
        &=\bm{c}_i^\top\mathbf{Pen}^{1/2}
        \left\{
            \mathbf{Z}\left(\mathbf{Z}+\Delta t\theta_0\mathbf{I}_L\right)^{-1}-\mathbf{I}_L
        \right\}
        \mathbf{Z}
        \left\{
            (\mathbf{Z}+\Delta t\theta_0\mathbf{I}_L)^{-1}\mathbf{Z}-\mathbf{I}_L
        \right\}
        \mathbf{Pen}^{1/2}\bm{c}_i
        +o_p(1)
        \\
        &=(\Delta t)^2\theta_0^2\bm{c}_i^\top\mathbf{Pen}^{1/2}
        \left(\mathbf{Z}+\Delta t\theta_0\mathbf{I}_L\right)^{-1}
        \mathbf{Z}
        \left(\mathbf{Z}+\Delta t\theta_0\mathbf{I}_L\right)^{-1}
        \mathbf{Pen}^{1/2}\bm{c}_i
        +o_p(1)
        \\
        &=\sum_{l=1}^L
        \frac{\tau_l}{(\theta_0^{-1}\Delta t^{-1}\tau_l+1)^2}
        (\bm{c}_i^\top\mathbf{Pen}^{1/2}\bm{e}_l)^2
        +o_p(1)
        \\
        &\leq \Delta t\theta_0\sum_{l=1}^{L_1-1}
        \tau_l(\bm{c}_i^\top\mathbf{Pen}^{1/2}\bm{e}_l)^2
        +\sum_{l=L_1}^L
        \tau_l(\bm{c}_i^\top\mathbf{Pen}^{1/2}\bm{e}_l)^2
        +o_p(1)
        \\
        &=o_p(1)
        \quad\text{as}\quad M\to\infty,
    \end{align*}
    where $L_1\in\mathbb{Z}^+$ is so defined that
    $\tau_{L_1}=\max\{\tau_L, (\Delta t\theta_0)^{1/2}\}$
    and diverges as $M\to\infty$ owing to \ref{cond:eigenvalue}.
\end{proof}

\begin{proof}{Proof of \autoref{prop:maximizer.T.hat}}{}
    Recall $\mathcal{P}_{BS_L}$ defined in the proof of \autoref{prop:x.i}.
    Writing $\hat{w}_{j,\alpha}=\mathcal{P}_{BS_L}\hat{w}_{j,\alpha}
        +(\mathcal{I}-\mathcal{P}_{BS_L})\hat{w}_{j,\alpha}$,
    with identity operator $\mathcal{I}$,
    one has $0<\|\mathcal{P}_{BS_L}\hat{w}_{j,\alpha}\|\leq 1$
    and $\int_{\mathcal{T}}\hat{X}_i\mathcal{P}_{BS_L}\hat{w}_{j,\alpha}
    =\int_{\mathcal{T}}\hat{X}_i\hat{w}_{j,\alpha}$ since $\hat{X}_i \in BS_L$ for all $i$.
    If $0<\|\mathcal{P}_{BS_L}\hat{w}_{j,\alpha}\|<1$
    (i.e., $(\mathcal{I}-\mathcal{P}_{BS_L})\hat{w}_{j,\alpha}>0$),
    then $\mathcal{P}_{BS_L}\hat{w}_{j,\alpha}/\|\mathcal{P}_{BS_L}\hat{w}_{j,\alpha}\|$
    satisfies that
    $$
        \hat{T}_{j,\alpha}\left(
            \frac{\mathcal{P}_{BS_L}\hat{w}_{j,\alpha}}
                {\|\mathcal{P}_{BS_L}\hat{w}_{j,\alpha}\|}
        \right)
        =\|\mathcal{P}_{BS_L}\hat{w}_{j,\alpha}\|^{2\alpha/(\alpha-1)}
        \hat{T}_{j,\alpha}(\hat{w}_{j,\alpha})
        >\hat{T}_{\alpha}(\hat{w}_{j,\alpha}),
    $$
    which violates the definition of $\hat{w}_{j,\alpha}$ \eqref{eq:w.j.hat}.
    This contradiction implies that
    $(\mathcal{I}-\mathcal{P}_{BS_L})\hat{w}_{j,\alpha}$ must be 0.
\end{proof}

\begin{proof}{Proof of \autoref{prop:ccc.converge}}{}
    Under conditions \ref{cond:x.derivative}--\ref{cond:unique},
    fixing $p$,
    \autoref{prop:x.i} and \citet[][ Theorem 1]{Zhou2019}
    assure us of the zero-convergence (in probability) of
    $\|\hat{\beta}_{p,\alpha}-\beta_{p,\alpha}\|$ as $N$ diverges.
    Since \autoref{prop:x.i} applies to $X^*$,
    the convergence of empirical classifiers follows.
\end{proof}
\end{appendix}


\CJShistory
\end{document}